\documentclass[10pt]{article}
\usepackage{times}
\usepackage{graphicx}

\title{An unusual stellar death on Christmas Day}

\author{C. C. Th\"one$^{1,2,*}$, A. de Ugarte Postigo$^{3}$, C. L. Fryer$^{4}$, K. L. Page$^{5}$, J. Gorosabel$^{1}$, M. A. Aloy$^{6}$, \\D. A. Perley$^{7}$, C. Kouveliotou$^{8}$, H. T. Janka$^{9}$, P. Mimica$^{6}$, J. L. Racusin$^{10}$,  H. Krimm$^{10,11,12}$, \\J. Cummings$^{10}$, S. R. Oates$^{13}$, S. T. Holland$^{10,11,12}$, M. H. Siegel$^{14}$, M. De Pasquale$^{13}$,\\ E. Sonbas$^{10,11,15}$, M. Im$^{16}$, W.-K. Park$^{16}$, D. A. Kann$^{17}$, S. Guziy$^{1,18}$, L. Hern\'andez Garc\'ia$^{1}$,\\ A. Llorente$^{19}$, K. Bundy$^{7}$, C. Choi$^{16}$, H. Jeong$^{19}$, H. Korhonen$^{21,22}$, P. Kubanek$^{1,23}$, J. Lim$^{24}$, \\A. Moskvitin$^{25}$, T. Mu\~noz-Darias$^{26}$, S. Pak$^{19}$, I. Parrish$^{7}$ \\
\small{$^{1}$ IAA - CSIC, Glorieta de la Astronom\'ia s/n, 18008 Granada, Spain}\\
\small{$^{2}$ Niels Bohr International Academy, Niels Bohr Institute, Blegdamsvej 17, 2100 Copenhagen, Denmark}\\
\small{$^{3}$ Dark Cosmology Centre, Niels Bohr Institute, Univ. of Copenhagen,}\\
\small{Juliane Maries Vej 30, 2100 Copenhagen, Denmark}\\
\small{$^{4}$ Los Alamos National Laboratory, MS D409, CCS-2, Los Alamos, NM 87545, USA}\\
\small{$^{5}$ Department of Physics \& Astronomy, Univ. of Leicester, University Road, Leicester LE1 7RH, UK}\\
\small{$^{6}$ Departamento de Astronomia y Astrofisica, Universidad de Valencia, 46100 Burjassot, Spain}\\
\small{$^{7}$ UC Berkeley, Astronomy Department, 601 Campbell Hall, Berkeley CA 94720, USA}\\
\small{$^{8}$ Space Science Office, VP62, NASA/Marshall Space Flight Center,
Huntsville, AL 35812, USA}\\
\small{$^{9}$ Max-Planck-Institut f\"ur Astrophysik, Karl-Schwarzschild-Str. 1, 85748 Garching, Germany}\\
\small{$^{10}$ NASA, Goddard Space Flight Center, Greenbelt, MD 20771, USA}\\
\small{$^{11}$ Universities Space Research Association, 10211 Wincopin Circle, Suite 500, Columbia, MD 21044-3432, USA}\\
\small{$^{12}$ Center for Research and Exploration in Space Science and Technology (CRESST)}\\
\small{$^{13}$ Mullard Space Science Laboratory, Holmbury St. Mary, Dorking, Surrey RH5 6NT, UK}\\ 
\small{$^{14}$ Dep. of Astronomy \& Astrophysics, Pennsylvania State Univ., 104 Davey Laboratory, University Park, PA 16802, USA}\\
\small{$^{15}$ University of Ad{\i}yaman, Department of Physics, 02040 Ad{\i}yaman, Turkey}\\
\small{$^{16}$ Center for the Exploration of the Origin of the Universe, Dept. of Physics \& Astronomy,}\\
\small{Seoul National University, 56-1 San, Shillim-dong, Kwanak-gu, Seoul, Korea}\\
\small{$^{17}$ Th\"uringer Landessternwarte Tautenburg, Sternwarte 5, 07778 Tautenburg, Germany}\\
\small{$^{18}$ Nikolaev National University, Nikolska 24, Nikolaev, 54030, Ukraine}\\
\small{$^{19}$ School of Space Research, Kyung Hee University, 1 Seocheon-dong, Giheung-gu,}\\
\small{Yongin-si, Gyeonggi-do 446-701, Korea}\\
\small{$^{20}$ Herschel Science Operations Centre, INSA, ESAC, Villafranca del
Castillo, PO Box 50727, I-28080 Madrid, Spain}\\
\small{$^{21}$ Finnish Centre for Astronomy with ESO (FINCA), University of Turku, V\"a{}is\"al\"antie 20, 21500 Piikki\"o, Finland}\\
\small{$^{22}$ Niels Bohr Institute, University of Copenhagen, Juliane Maries Vej 30, 2100 Copenhagen, Denmark}\\
\small{$^{23}$ Institute of Physics, Na Slovance 2, 180 00, Prague 8, Czech Republic}\\
\small{$^{24}$ Dept. of Astronomy and Space Science, Kyung Hee University, 1 Seocheon-dong,}\\
\small{Giheung-gu, Yongin-si, Gyeonggi-do 446-701, Korea}\\
\small{$^{25}$ Special Astrophysical Observatory of the Russian Academy of Sciences, Nizhnij Arkhyz 369167, Russia}\\
\small{$^{26}$ INAF - Osservatorio Astronomico di Brera, Via E. Bianchi 46, 23807 Merate, Italy}\\
\small{$^\ast$To whom correspondence should be addressed; E-mail: cthoene@iaa.es}
}

\date{}

\begin{document} 

\baselineskip24pt

\maketitle 

\newpage

{\bf 
\noindent
Long Gamma-Ray Bursts (GRBs) are the most dramatic examples of massive stellar deaths, usually associated with supernovae \cite{WB}. They release ultra-relativistic jets producing non-thermal emission through synchrotron radiation as they interact with the surrounding medium \cite{ZhangAG}. Here we report observations of the peculiar GRB 101225A (the ``Christmas burst''). Its $\gamma$-ray emission was exceptionally long and followed by a bright X-ray transient with a hot thermal component and an unusual optical counterpart. During the first 10 days, the optical emission evolved as an expanding, cooling blackbody after which an additional component, consistent with a faint supernova, emerged. We determine its distance to 1.6 Gpc by fitting the spectral-energy distribution and light curve of the optical emission with a GRB-supernova template. Deep optical observations may have revealed a faint, unresolved host galaxy. Our proposed progenitor is a helium star-neutron star merger that underwent a common envelope phase expelling its hydrogen envelope. The resulting explosion created a GRB-like jet which gets thermalized by interacting with the dense, previously ejected material and thus creating the observed black-body, until finally the emission from the supernova dominated. An alternative explanation is a minor body falling onto a neutron star in the Galaxy \cite{SergioNat}.
}
\vspace{5mm}
 
\noindent
On Dec. 25, 2010, 18:37:45 UT (T$_{0}$), the {\it Swift} Burst Alert Telescope (BAT, 15-350 keV) detected GRB 101225A, one of the longest GRBs ever observed by {\it Swift} \cite{060218Swift} with T$_{90}>2000$ s (the time in which 90\% of the $\gamma$-ray energy is released, \cite{Kouveliotou93}). A bright X-ray afterglow was detected for two days and a counterpart in ultraviolet, optical and infrared bands could be observed from 0.38\,h to two months after the event (see the Supplementary Information, SI). No counterpart was detected at radio frequencies \cite{radioGCN1, radioGCN2}.

The most surprising feature of GRB 101225A is the spectral energy distribution (SED) of its afterglow. The X-ray SED is best modeled with a combination of an absorbed power-law and a black-body (BB). The UV/optical/NIR (UVOIR) SED (see Fig. 1) can be modeled with a cooling and expanding BB model until 10 days post burst, after which we observe an additional spectral component accompanied by a flattening of the light curve (Fig. 2). This behaviour differs from a normal GRB where the SED follows a power-law due to synchrotron emission created in shocks when the jet hits the interstellar medium (e.g., \cite{ZhangAG}). 

An optical spectrum taken two nights after the burst does not show any spectral lines (see SI). Were fit the SED and light curve with the template of SN 1998bw, a SN Type Ic associated with GRB 980425 \cite{Galama98}, and obtain a redshift of $z=0.33$. At this distance, the SN has an absolute peak magnitude of only M$_\mathrm{V}=-16.7$ mag, which makes it the faintest SN associated with a long GRB \cite{Soderberg06, Wiersema08}. In contrast, the $\gamma$-ray isotropic-equivalent energy release at $z=0.33$ is $>1.4\times10^{51}$ erg, typical for other long GRBs but more luminous than most other low-redshift GRBs associated with SNe \cite{Kann10}. We detect a possible host galaxy in $g^\prime$ and $r^\prime$ bands with OSIRIS/GTC at 6 months after the burst with an absolute magnitude of only M$_{g\mathrm{,abs}}=-13.75$ mag, $\sim$\,2\,mag fainter than any other GRB host \cite{Wiersema07}. Although its blue colour matches that of a star-forming galaxy, our observations do not allow us to resolve it as an extended source.

At $z=0.33$, the X-ray BB has a radius of $\sim2\times10^{11}$\,cm ($\sim$1 R$_\odot$) and a temperature of $\sim1$ keV ($10^7$\,K) at 0.07\,d with little temporal evolution. Such a thermal component, attributed to the shock breakout from the star, had also been observed for XRF 060218 \cite{Campana06}, XRF 100316D/SN 2010dh \cite{Starling11} and GRB 090618 \cite{Page11}, all nearby GRBs associated to Type Ic SNe \cite{Pian06,Mazzali06,Starling11,Cano11}, with similar temperatures but larger radii. The UVOIR BB starts with a radius of $2\times10^{14}$ cm ($\sim$13 AU) and a temperature of $8.5\times10^4$ K at similar times 
and evolves considerably over the next 10 days reaching a radius of 7$\times$10$^{14}$cm and temperature of 5,000\,K. The evolution of the two BB components suggests that they must stem from different processes and regions.

An appealing model is a Helium - neutron star (NS) merger with a common envelope (CE) phase, a model that had been proposed earlier as a possible progenitor for GRBs \cite{Fryer98, Zhang01, Barkov10}.  In this scenario, a binary system consisting of two massive stars survives the collapse of the more massive component to a neutron star (NS). When the second star leaves the main sequence and expands, it engulfs the NS, leading to a CE phase and the ejection of the hydrogen envelope and part of the helium core as the remnant spirals into the center of the second star. When the NS reaches the center, angular momentum forms a disk around the remnant of the merger, allowing for the formation of a GRB-like jet. The remnant might be a magnetar whose prolonged activity can explain the very long duration of the GRB.  

The interaction of this ultra-relativistic, well-collimated jet with the previously ejected CE material can explain both the X-ray and UVOIR emission components. Estimating that the inspiral takes 5 orbits or 1.5 yr and material is ejected at escape velocity, the outer ejecta are at a few $10^{14}$ cm at the time of the merger, consistent with the radius of the UVOIR BB. We assume that the ejecta form a broad torus with a narrow, low-density funnel along the rotation axis of the system that permits the passage of the $\gamma$-radiation generated in the jet. Most of the jet hits the inner boundary of the CE-ejecta and only a small fraction of it propagates through the funnel. The X-ray emission is produced by shocks created by the interaction of the jet with the inner boundary of the CE shell.  As the jet passes through the funnel, it decelerates due to the increased baryon load and shear with the funnel walls so that no regular afterglow signature is produced. 
When the now mildly relativistic, mass-loaded jet breaks out of the CE-ejecta, it produces the UVOIR emission in the first 10\,d. As the supernova shock expands beyond the CE shell we observe a small bump in the lightcurve at $\sim$30\,d. The He-NS merger scenario naturally assumes a relatively small Ni-production, leading to a weak SN. 


A somewhat similar scenario might also explain XRF 060218 albeit with a different progenitor system producing a brighter SN and a fainter GRB. The remnant might be a magnetar whose prolonged activity can explain the very long duration of the GRB. A thermal component in the optical can also be fitted for the X-ray outburst associated with SN 2008D, a Type Ib SN in NGC 2770 \cite{Soderberg08}, which, however, had no thermal component in X-rays. And last, it is possible that a GRBs shows a thermal component in X-rays (e.g., GRB 090618) but shows a classical, bright power-law shaped afterglow \cite{Cano11}. GRB 101225A might be a member of a newly defined class of ``blackbody-dominated'', SN-associated long-duration GRBs arising in very dense environments created by the progenitor systems themselves, which thermalizes the high-energy output from the collapsing star. The non-relativistic, uncollimated emission from this scenario makes it difficult to detect them at higher redshifts. This makes GRB 101225A a fortunate coincidence that allows us to derive conclusions about the progenitor system and its environment by a new variety of massive stellar death, which had so far been only proposed to exist theoretically. 

\newpage

\noindent
{\bf Acknowledgements}\\
Based on observations collected at CAHA/Calar Alto, GTC/La Palma, the Liverpool Telescope at ORM/La Palma, the McDonald Observatory at the University of Texas at Austin, Gemini-North and Keck on Big Island/Hawaii. We thank J. S. Bloom for helping with the Keck observations.
The Dark Cosmology Centre is funded by the DNRF. KLP, SRO and MDP acknowledge the support of the UK Space Agency. JG, AJCT, SG, PK, MAA and PM are partially supported by MICINN. HTJ acknowledges support by a DFG grant. MI, WKP, CC, JL and SP acknowledge support from CRI/NRF/MEST of Korea. AM acknowledges support from the Russian government. The first author acknowledges a personal connection to the event 31 years prior to the burst.
\ \\[5mm]
{\bf Author contributions}\\
CT did the overall management of the observations and modeling, the analysis of the spectra and wrote most of the manuscript. ADUP did the UVOIR BB modeling, SN template fitting,  most of the optical/IR photometry and lead the GTC observations. CF suggested and investigated the progenitor system. KLP did the X-ray analysis, JG worked on the SN templates and the photometric calibrations for the optical/IR data. MAA did the modeling of the UVOIR BB and X-ray emission from numerical simulations. DP contributed with the observation and analysis of the late Gemini and Keck data. CK investigated possible progenitor models. HTJ, PM and AL contributed to the theoretical modeling. JLR, HK, JC, SRO, STH, MHS, MDP and ES did the analysis of the Swift data. MI, WKP, CC, HJ, JL and SP contributed the McD 2.1m data, AM the late BTA 6m data, KB and IP the late Keck spectrum. DAK did the comparison of SN stretching factors and luminosities. SG and LHG helped with the optical photometry. HK and TMD investigated alternative interpretations of the event, PK assisted with the manuscript.

\newpage

\begin{figure}[ht!]
   \centering
   \includegraphics[width=13.5cm]{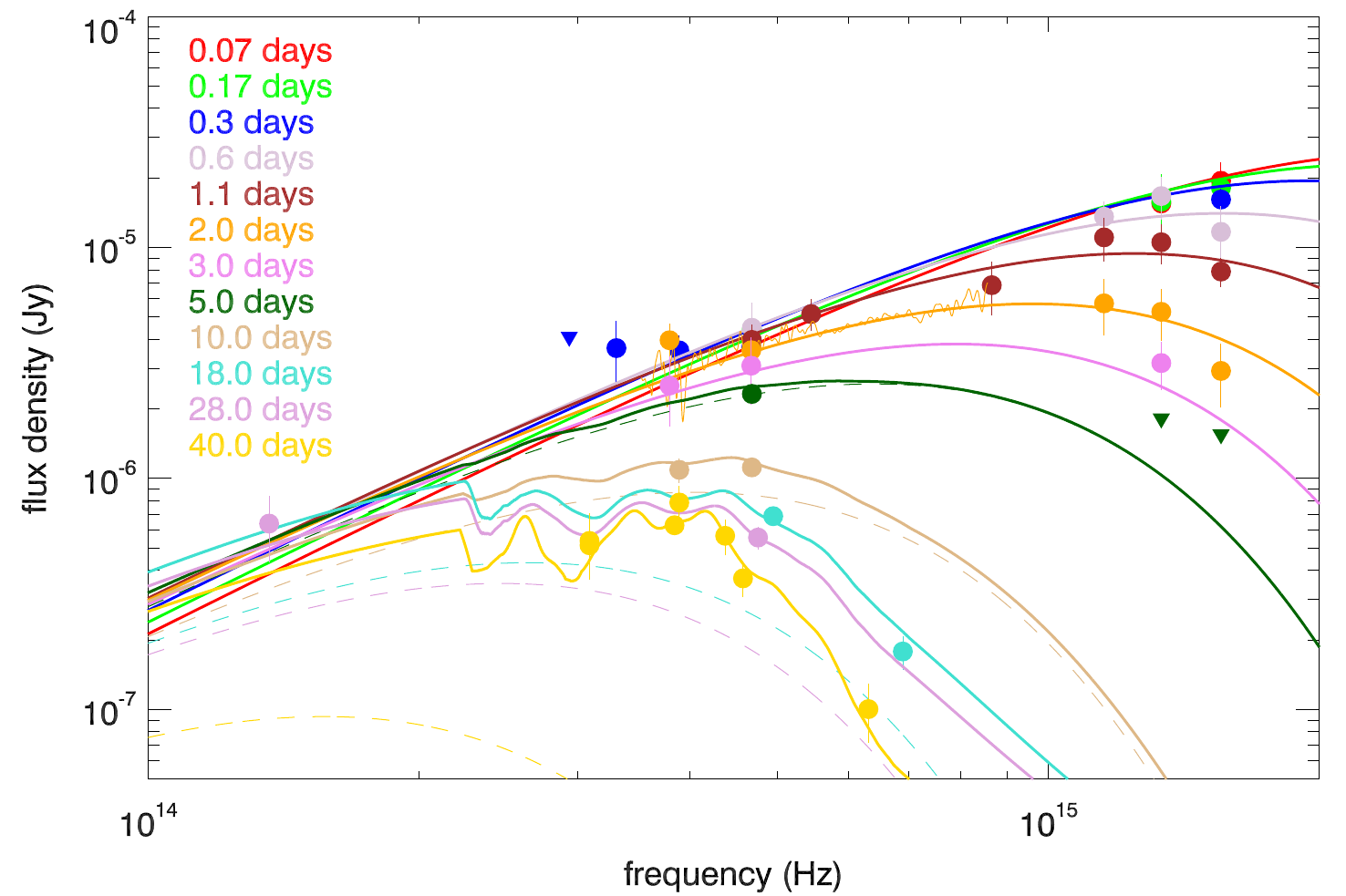}
          
         \label{Fig:sed}
   \end{figure}
\noindent   
{\bf\small Figure 1: Temporal evolution of the UVOIR spectral energy distribution}. {\small The UV, optical and IR counterpart were detected by UVOT, the UV telescope onboard {\it Swift} and several ground based facilities from 0.38h to nearly 2 month after the GRB. This plot shows the evolution of the SED from the onset of the optical observations at 0.07 days to 40 days for all epochs with sufficient data to model the SED shape. Filled circles are detections, triangles mark upper limits. The orange line on top of the BB model at 2.0 days shows our flux-calibrated spectrum taken with the OSIRIS/GTC. The SED evolution requires two different components, a simple expanding and cooling BB up to 10\, days and an additional SN component for the last 3 epochs. The solid line shows the combined evolution taking both the BB and the SN beyond 5 days into account, whereas the dashed line shows the evolution of the BB component alone. The UVOIR BB evolves from an initial temperature of 43,000 K (0.07 d) to 5,000 K (18 d) and increases in radius from 2$\times$10$^{14}$ cm to 7$\times$10$^{14}$ cm at the same timescale. The SN component is best fit with a template of the broad-line Type Ic SN 1998bw which was associated to GRB 980425. For the fit of the SN component we used the SED at 40 days. Reanalyzing UVOIR data of XRF 060218 \cite{Campana06} and SN 2008D \cite{Soderberg08} we find a similar thermal component over the first 3--4 days after which the SN starts to dominate. The UVOIR BB of XRT 060218 shows a similar radius, temperature and evolution as GRB 101225A and is equally inconsistent with radius and temperature from the X-ray BB. SN 2008D shows a steeper evolution in the radius, consistent with the continuation from the shock breakout, a thermal X-ray component was not observed (see SI). }

\newpage

\begin{figure}[ht!]
  \centering
  \includegraphics[width=15cm]{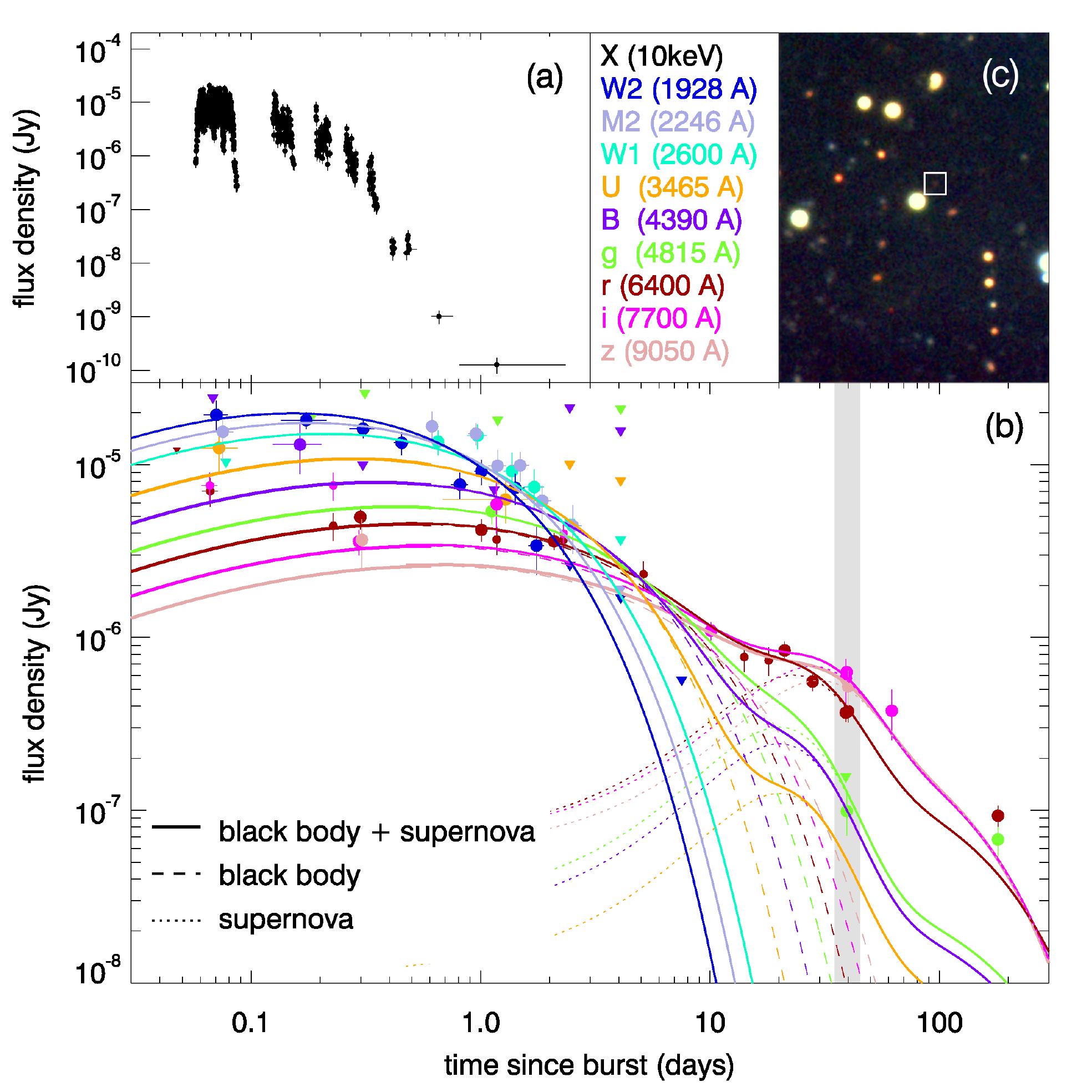}
        \label{Fig:lc}
   \end{figure}

\noindent 
{\bf\small Figure 2: Light curves of GRB 101225A in X-ray and UV/optical/IR bands}. {\small (a) X-ray lightcurve, (b) light curves in UV to IR bands. Filled points are detections, triangles upper limits. The solid lines denote the combined light curve from the BB and the SN component, excluding the contribution from the host galaxy. The evolution of the BB component is shown as a dashed line, the SN as a dotted line. Observations started almost simultaneously in X-rays and optical/UV wavelengths. The X-rays reached a peak flux of $4.34\times10^{-9}$ erg cm$^{-2}$ s$^{-1}$. After an initial shallow decay of slope $t^{-1.108\pm0.011}$ up to 21 ks, the X-rays show a strong decay with a slope of $t^{-5.95\pm0.20}$, inconsistent with synchrotron emission. The UVOIR light curves have a shallow maximum at the beginning, with different peak times for the individual bands due to the maximum of the BB emission passing through the spectrum. The second component emerging at around 10 days post-burst is the contribution of an underlying SN, modeled with the GRB-SN 1998bw as a template, stretched in time by a factor of 1.25 and decreased in luminosity by a factor of 12 (in restframe). The absolute luminosity of the SN is M$_V=-16.7$ mag, the faintest SN associated with a GRB. At $\sim$\,180\,days we detect the very faint host at magnitudes of $g^\prime=27.36\pm0.27$ and $r^\prime=26.90\pm0.14$ or M$_\mathrm{abs}=-13.7$ (0.001\,L*). (c) Color image of the field of GRB 101225A observed at 40 days (indicated by a grey bar in panel b) with the afterglow marked by a box. }

\newpage

\begin{center}
{\huge Supplementary information}
\end{center}

\vspace{2cm}
\section{BAT data analysis and fitting}

\begin{figure}[!ht]
\begin{center}
\includegraphics[width=12cm]{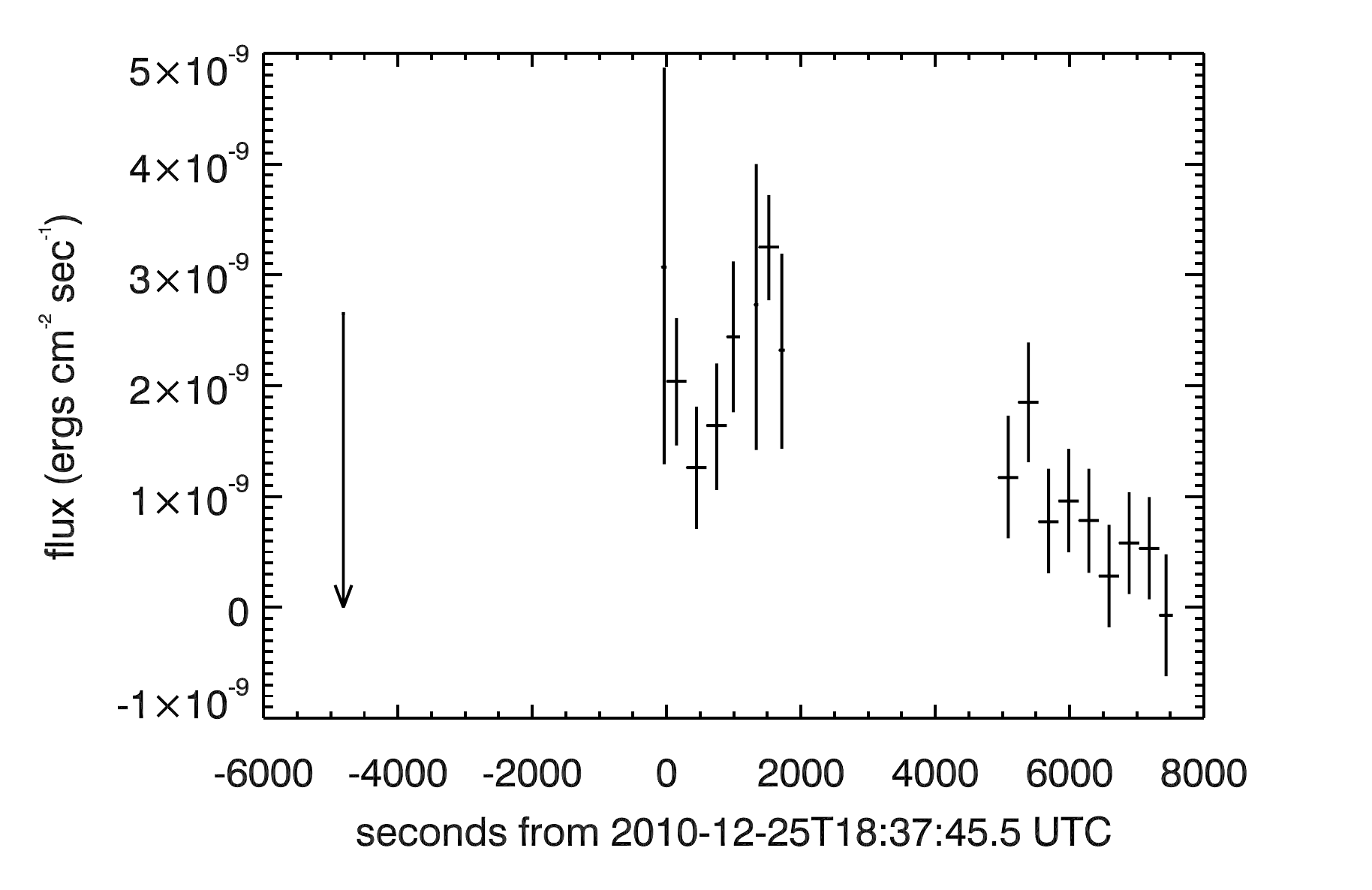}
\caption{{\bf BAT $\gamma$-ray lightcurve in the 15-150 keV band.} The sum of the flux was obtained using a fixed power-law index of $\Gamma=1.87$ from a fit to the most intense part of the burst. The burst started before the beginning of the BAT data at $\sim \mathrm{T-T_0}=-100$\,s and probably continued while the source was not in the BAT field of view from $\mathrm{T-T_0}=+1091$ to $\mathrm{T-T_0}=+1372$\,s. The latest upper limit before the burst was $2.65\times10^{-9}$ erg cm$^{-2}$ s$^{-1}$ at $\mathrm{T-T_0}=-4950$ s. Error bars are at 90\% confidence.}
\label{101225:BAT}
\end{center}
\end{figure}

GRB\,101225A was detected by BAT on-board the {\it Swift} satellite \cite{Gehrels04} on Dec. 25, 2010 at $\mathrm{T_0}=18$:37:45 UT as an image trigger \cite{Racusin10}. It was already in progress both when the source entered the BAT field of view and when it left the field of view due to {\it Swift} orbit-constrained slews \cite{Palmer10, Cummings10}. Therefore, we can only give lower limits on the total burst fluence and the $\mathrm{T}_{90}$ duration. The total fluence of the intervals covered in the observations adds up to $(5.6\pm0.7)\times$ 10$^{-6}$ erg\,cm$^{-2}$ (implying a total energy release in $\gamma$-rays of E$_{\gamma,iso}>1.4\times10^{51}$ erg at $z=0.33$.), which is a lower limit to the total gamma-ray emission. No emission was detected in a previous observation of the field at $\mathrm{T-T_0}=-4950$ s and we therefore put a lower limit on the duration of $\mathrm{T}_{90}>2000$ s. This is one of the highest durations ever observed for a GRB, comparable to the longest burst observed by {\it Swift}, GRB\,090417B \cite{Holland10}. The BAT light curve is shown in Fig. \ref{101225:BAT}. The BAT-observed peak flux of (3.25$\pm$0.47) $\times$ 10$^{-9}$ erg cm$^{-2}$ s$^{-1}$ in the $15-150$ keV range occurred in the interval $\mathrm{T-T_0}=+1372$ to $\mathrm{T-T_0}=+1672$ s. No other $\gamma$-ray instrument detected GRB\,101225A, although the MAXI instrument on board the ISS ($2-10$ keV) reported a marginal detection at $\mathrm{T-T_0}=+1002$ s coincident with the BAT position \cite{Serino10}.

The time-averaged spectra from $\mathrm{T_0}$ to $\mathrm{T-T_0}=+963$ s and from $\mathrm{T-T_0}=+1372$ to $\mathrm{T-T_0}=+1672$ s are best fit by simple power-law models with photon indices of $\Gamma=1.9\pm0.4$ and $1.9\pm0.2$, respectively. The total fluences in each time period in the $15-150$ keV band taking these fits are ($1.7\pm0.4$)$\times10^{-6}$ and ($9.0\pm0.2$) $\times10^{-7}$ erg cm$^{-2}$. All quoted errors are at the 90\% confidence level.
The BAT spectra are almost equally well parametrised by models using a cutoff power-law or a blackbody fit due to the low signal-to-noise ratio of the event. E$_\mathrm{peak}$ using a cutoff model is poorly constrained to $38\pm20$ keV. The blackbody fit gives a temperature of $\mathrm{kT}=10.1\pm1.1$ keV. Errors are at the 68\% confidence level. 

We also examined the BAT data to search for persistent emission after the trigger. For this we used the daily sky image mosaics produced as part of the BAT hard X-ray transient monitor which cover a single energy band of $15-50$ keV.  We found a $5.3\sigma$ excess ($0.0048\pm0.0009$ count cm$^{-2}$ s$^{-1}$) on Dec. 25, 2010 (MJD 55555), the day of the trigger, and a positive excess in the count rate ($\geq1\sigma$ or 0.0011 count cm$^{-2}$ s$^{-1}$) over the next ten days (until MJD 55565). We determine the probability that such a sequence of excess rates would occur by chance.  To do this, we examine the light curves of 106 ``blank sky'' points tracked in the BAT transient monitor.  These are points chosen randomly across the sky at least 10 arcmin from any known X-ray source.  Any positive flux from these points is expected to be due to chance fluctuations. In these 106 light curves ($>200,000$ data points), we find only one sequence of six consecutive days showing a positive excess and none with more than six days. The chance probability of ten days of excess flux is less than 1/200,000, so the observed prolonged emission is likely real.

\section{XRT data analysis and fitting}

The {\it Swift} XRT data were processed with version 3.7 of the XRT data reduction software (released as part of {\tt HEASoft 6.10} on 2010-09-28) and the corresponding calibration files used for subsequent spectral analysis. The object was detected by XRT from 1.4 ks to 10$^5$ s after the trigger.  Data were collected in Windowed Timing (WT) mode for the first 7.3 ks after the trigger followed by Photon Counting (PC) mode for the rest of the observations. The peak flux in X-rays is $4.3\times10^{-9}$ erg cm$^{-2}$ s$^{-1}$, the total observed fluence $8.2\times10^{-6}$ erg cm$^{-2}$ and the unabsorbed fluence $1.1\times10^{-5}$ erg cm$^{-2}$. At $z=0.33$ this corresponds to a total energy release in X-rays of $3.6\times10^{51}$ erg. Spectra were extracted for individual snapshots of data (one snapshot corresponds roughly to one orbit constrained by the observability of the object during the orbit) and were further timesliced into 100 s bins for the initial snapshot ($1.4-1.8$\,ks after the trigger). 

\begin{figure}[!ht]
\begin{center}
\includegraphics[width=12cm]{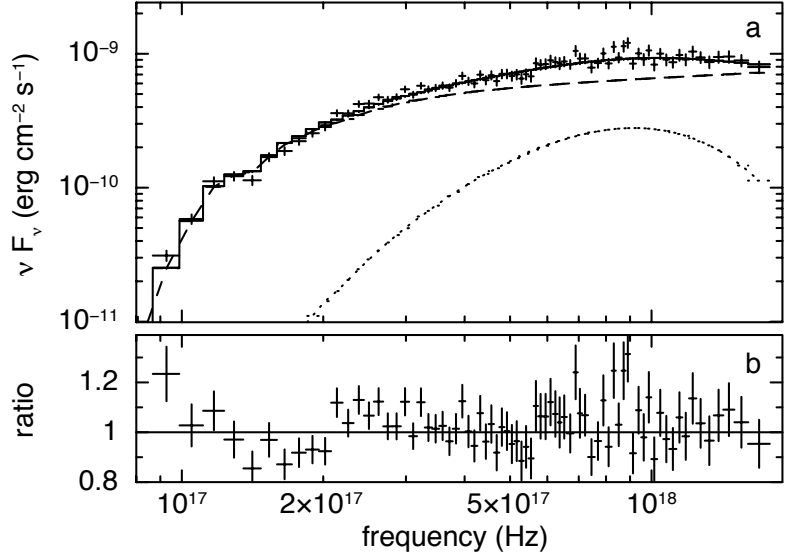}
\caption{{\bf Fit to the X-ray spectrum.} Panel a): X-ray spectrum from XRT during the first snapshot, the dashed line indicates the contribution of the power-law component, the dotted line shows the BB component. Panel b): Ratio between the observed data and the fitted model.}
\label{101225:Xrayspec}
\end{center}
\end{figure}

\begin{figure}[!ht]
\begin{center}
\includegraphics[width=10cm]{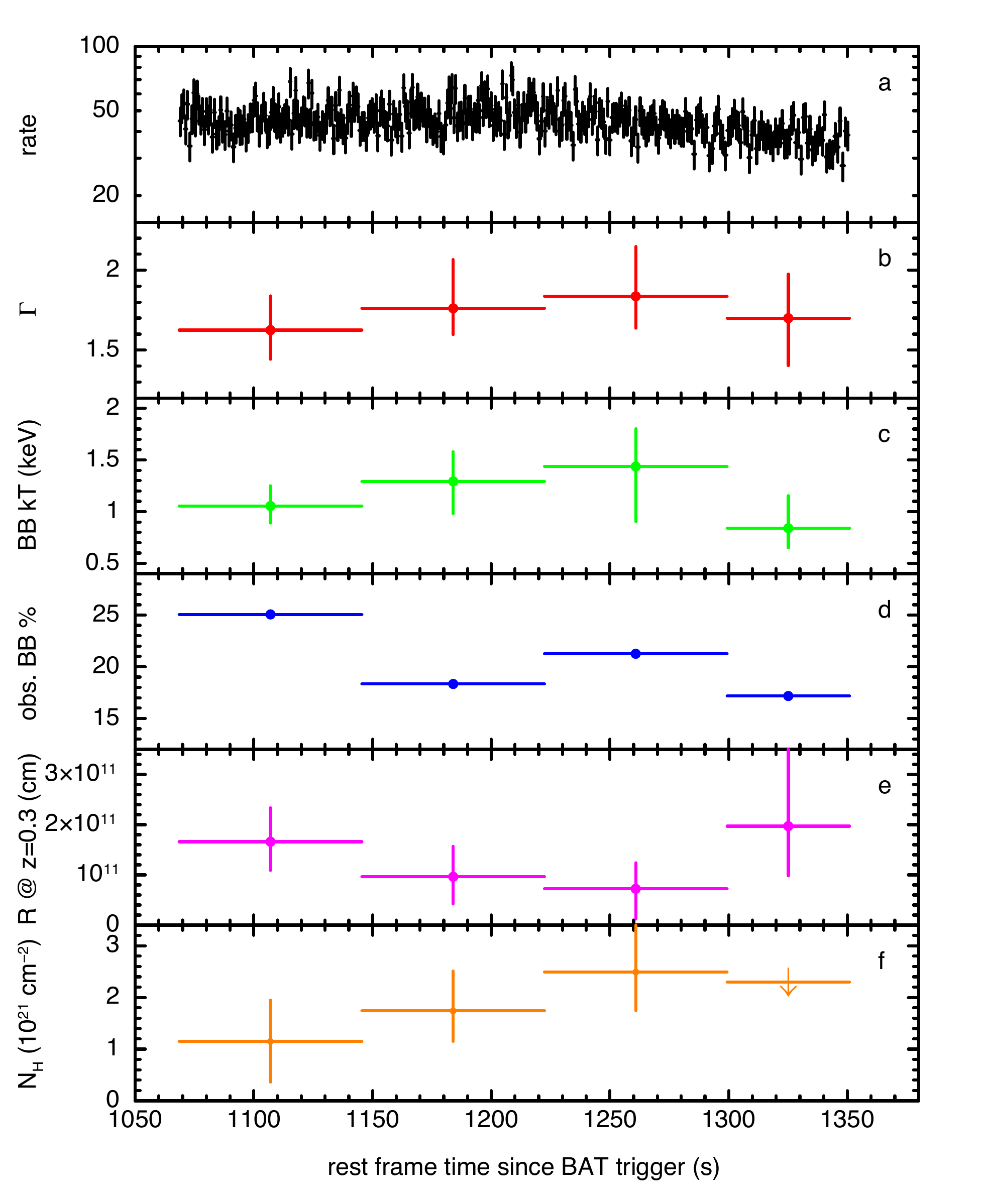}
\caption{{\bf Results from the fits to the first snapshot of the X-ray data.} The panels show from top to bottom: a) count rate during the first snapshot, b) photon index $\Gamma$, c) BB Temperature in keV, d) the contribution of the BB to the total emission in percent, e) radius of the emitting BB at $z=0.33$ and f) total absorbing column density in X-rays (the Galactic column density in the line-of-sight is $7.9 \times 10^{20}$ cm$^{-2}$).}
\label{101225:Xrayanalysis}
\end{center}
\end{figure}

We tried a variety of fits to the X-ray data, using {\tt XSPEC version 12.6.0}, with the result that an absorbed power-law plus BB component provided a good fit to the data (see Fig.~\ref{101225:Xrayspec}). For the fit we used the T\"ubingen-Boulder absorption model with the Wilms abundances \cite{Wilms00} and Verner absorption cross-section \cite{Verner96}. As shown in Fig. \ref{101225:Xrayanalysis}, there is little spectral evolution within the first snapshot of data. The best fit taking the entire first 367 s of data is a power-law with a photon index $\Gamma=1.83^{+0.13}_{-0.10}$, a blackbody of temperature $0.96\pm0.13$ keV ($1.11\times10^{7}$ K) and a total absorbing column of ($2.2\pm0.3$)$\times10^{21}$ cm$^{-2}$, for a $\chi^2$ of 420.7 for 379 degrees of freedom. The Galactic column density in this direction is $7.9\times10^{20}$ cm$^{-2}$. 

The inclusion of the BB is significant at the $>99.9999$\% level, the contribution of the BB to the total emission is around 20\% (see Fig. \ref{101225:Xrayanalysis}). The second snapshot of data (also in WT mode) is again better fit with a BB in addition to the power-law, with $\Gamma=2.18^{+0.12}_{-0.09}$, black-body $\mathrm{kT}=0.99^{+0.15}_{-0.17}$ keV and $\mathrm{NH}=(2.7\pm0.2)\times10^{21}$ cm$^{-2}$, with $\chi^2$/d.o.f = 378/421. This BB is significant at 99.987\%. For the X-ray data after the second snapshot, no BB component is required and a simple absorbed power-law provides an acceptable fit, likely due to the lower signal-to-noise ratio at later times.

\begin{table*}[!ht]
\caption{Result of different model fits to the X-ray data of the first snapshot. PL is a pure power-law model, PL+BB a combination of a simple PL and a BB with one temperature while a PL+diskBB includes emission from BBs with different temperatures, PL+compt includes a comptonized component in addition to the power-law. $\Gamma$ is the photon index of the power-law, kT the BB temperature, the column density is the total density, including the Galactic absorption. The last column shows the $\chi^2$ of the different fits and the F-test value compared to the simple absorbed PL model. \label{table:xrayfits}}
\begin{center}
\begin{tabular}{lllll} \hline
Model  &$\Gamma$&kT (keV)&N$_\mathrm{H}$ (10$^{22}$ cm$^{-2}$)    &    $\chi^2$/d.o.f. (F-test)\\ \hline\hline
PL     &         1.72$\pm$0.03        &     &        0.24$\pm$0.02   &  468/381\\
PL+BB    &       1.83$^{+0.13}_{-0.10}$& 0.96$\pm$0.13   &  0.22$\pm$0.03  &   421/379 (1.95$\times 10^{-9}$)\\
PL+diskBB   &    1.79$^{+0.36}_{-0.22}$ &1.64$\pm$0.35 &    0.18$\pm$0.04   &  417/379 (3.19$\times 10^{-10}$)\\
PL+compt   &     1.79$\pm$0.04  &  $<$22      &       0.31$^{+0.03}_{-0.02}$ &424/377 (1.55$\times 10^{-7}$)\\ \hline
\end{tabular}
\end{center}
\end{table*}

\begin{figure}[!ht]
\begin{center}
\includegraphics[width=12cm]{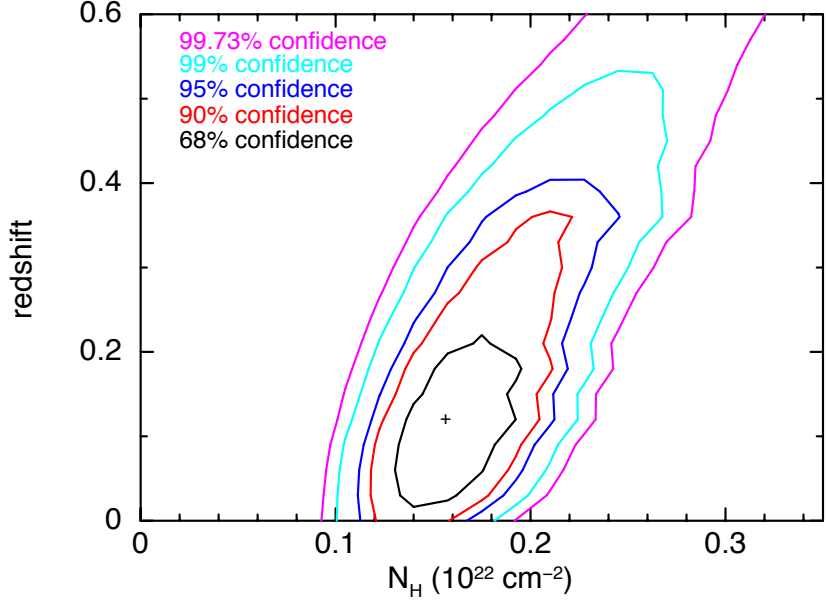}
\caption{{\bf Redshift vs. absorbing column density confidence contours.} The fit is based on the data from the first snapshot. Plotted are the 1, 2 and 3 $\sigma$ contours together with the 90\% and 99\% confidence interval.}
\label{fig:Xraycontours}
\end{center}
\end{figure}

The detection of N$_H$ in excess of the Galactic column density can be used to constrain the redshift since column density and redshift are coupled. We fitted the spectrum of the first orbit fixing the Galactic absorption to $7.93\times10^{20}$ cm$^{-2}$. In addition to the absorbed powerlaw plus BB component as described above we allowed for a third component leaving both the column density and the redshift free to vary (in contrast to the fit described above where the redshift was fixed to 0.3). The resulting contours are plotted in Fig. \ref{fig:Xraycontours}. The 99\% upper limit for the redshift is 0.5, the 90\% upper limit on the redshift is 0.35, consistent with our findings from the UVOIR SED fitting.
 
We also checked for a possible periodicity in the X-ray data. To this end, light curves were extracted with 18 ms bins which is the best time resolution available for WT mode. Using the {\tt Kronos powspec} tool, no significant periodic signal was identified with a frequency between 0.005 and 28 Hz (0.04 and 200 s) in either the first or second snapshot of data. The 3$\sigma$ limit on the variation in flux of any periodic signal during the first WT snapshot is 6.3\%.

\section{UV, optical and IR data analysis}

\subsection{{\it Swift}/UVOT}
{\it Swift}/UVOT began observing GRB\,101225A 1373 s after the BAT trigger, simultaneous with the XRT observations. The automatic target sequence did not start until the end of the BAT image trigger at $\sim23$ minutes. The optical counterpart was found to be blue, with strong detections in the UV filters ($uvw1$, $uvm2$, $uvw2$), weak detections in the $b$ and $u$ filters, and no detection in the $v$ filter. The data were processed using the standard {\it Swift} software tool {\tt uvotmaghist} within {\tt HEAsoft 6.9} and the latest calibration files (20101231). 

We extracted counts using a circular aperture with a radius of 5 arcsec for count rates above 0.5 counts s$^{-1}$, and a 3 arcsec aperture where the count rate drops below 0.5 counts s$^{-1}$, as well as a source-free background region. The tool {\tt uvotmaghist} applies coincidence-loss corrections and aperture corrections. The count rates were converted to flux density using standard photometric calibration \cite{Poole08, Breeveld10}.

\begin{table}[ht!]
{\tiny 
\caption{Log of observations in the UVOIR range. Values are not corrected for Galactic extinction.}             
\label{table:log}      
\begin{center}                        
\begin{tabular}{c c c c c c}       
\hline               
Mid t-t$_0$      	& Exposure 			&  Filter 		& Telescope			& Mag$_{AB}$  & Flux\\
(days)		&	(s)	    			&           		&                      			&                          & ($\mu$Jy) \\
\hline                        
Premaging	& 3$\times$500		& \textit{g$^\prime$}		& 3.5mCFHT			& $>$26.9 (27.2$\pm$0.5)	& $<$0.06 (0.048$\pm$0.22)	\\
Premaging	& 3$\times$500		& \textit{i$^\prime$}		& 3.5mCFHT			& $>$25.5 	& $<$0.22	\\
\hline\hline 
 0.01848 &    168& \textit{w2} & UVOT & $>$21.36 & $<$ 10.38\\
 0.07041 &   1431& \textit{w2} & UVOT & 21.56$\pm$ 0.20 &   8.65$\pm$ 1.80\\
 0.17373 &   6719& \textit{w2} & UVOT & 21.63$\pm$ 0.11 &   8.06$\pm$ 0.88\\
 0.30739 &   6679& \textit{w2} & UVOT & 21.76$\pm$ 0.12 &   7.19$\pm$ 0.85\\
 0.45280 &   5805& \textit{w2} & UVOT & 21.96$\pm$ 0.15 &   5.96$\pm$ 0.91\\
 0.81302 &  12039& \textit{w2} & UVOT & 22.57$\pm$ 0.17 &   3.42$\pm$ 0.58\\
 1.00869 &  11753& \textit{w2} & UVOT & 22.37$\pm$ 0.16 &   4.08$\pm$ 0.67\\
 1.41736 &  23440& \textit{w2} & UVOT & 22.61$\pm$ 0.20 &   3.27$\pm$ 0.66\\
 1.75211 &  23368& \textit{w2} & UVOT & 23.45$\pm$ 0.30 &   1.51$\pm$ 0.49\\
 2.44862 &  74516& \textit{w2} & UVOT & $>$23.73 & $<$  1.17\\
 4.07964 & 138747& \textit{w2} & UVOT & $>$24.20 & $<$  0.76\\
 7.52954 & 377712& \textit{w2} & UVOT & $>$25.39 & $<$  0.25\\
\hline
 0.01818 &    319& \textit{m2} & UVOT & $>$20.81 & $<$ 17.14\\
 0.07515 &   1431& \textit{m2} & UVOT & 21.97$\pm$ 0.31 &   5.89$\pm$ 1.94\\
 0.61452 &    899& \textit{m2} & UVOT & 21.90$\pm$ 0.21 &   6.34$\pm$ 1.33\\
 0.95369 &  12104& \textit{m2} & UVOT & 22.00$\pm$ 0.15 &   5.74$\pm$ 0.83\\
 1.18487 &  18396& \textit{m2} & UVOT & 22.47$\pm$ 0.23 &   3.74$\pm$ 0.87\\
 1.48860 &  23468& \textit{m2} & UVOT & 22.46$\pm$ 0.19 &   3.76$\pm$ 0.73\\
 1.85629 &  29528& \textit{m2} & UVOT & 22.97$\pm$ 0.22 &   2.35$\pm$ 0.53\\
 2.51507 &  40973& \textit{m2} & UVOT & 23.34$\pm$ 0.30 &   1.68$\pm$ 0.53\\
 4.08285 & 138701& \textit{m2} & UVOT & $>$24.25 & $<$  0.73\\
\hline
 0.01846 &    318& \textit{w1} & UVOT & $>$21.15 & $<$ 12.64\\
 0.07752 &   1431& \textit{w1} & UVOT & $>$22.10 & $<$  5.26\\
 0.65205 &   5571& \textit{w1} & UVOT & 21.81$\pm$ 0.17 &   6.88$\pm$ 1.15\\
 0.96984 &  16520& \textit{w1} & UVOT & 21.72$\pm$ 0.16 &   7.46$\pm$ 1.19\\
 1.37145 &  18904& \textit{w1} & UVOT & 22.23$\pm$ 0.26 &   4.65$\pm$ 1.25\\
 1.71425 &  29052& \textit{w1} & UVOT & 22.46$\pm$ 0.25 &   3.76$\pm$ 0.97\\
 2.44380 &  74329& \textit{w1} & UVOT & $>$22.97 & $<$  2.36\\
 4.07673 & 138760& \textit{w1} & UVOT & $>$23.22 & $<$  1.86\\
\hline
 0.01789 &    169& \textit{u} & UVOT & $>$20.46 & $<$ 23.68\\
 0.07228 &   2579& \textit{u} & UVOT & 21.59$\pm$ 0.28 &   8.42$\pm$ 2.47\\
 1.28374 & 103571& \textit{u} & UVOT & 22.33$\pm$ 0.26 &   4.24$\pm$ 1.14\\
 2.44533 &  74215& \textit{u} & UVOT & $>$21.82 & $<$  6.80\\
 4.07781 & 138647& \textit{u} & UVOT & $>$22.06 & $<$  5.46\\
\hline         
 0.01817 &    169& \textit{b} & UVOT & $>$19.94 & $<$ 38.37\\
 0.06802 &   1430& \textit{b} & UVOT & $>$20.86 & $<$ 16.51\\
 0.16319 &   6726& \textit{b} & UVOT & 21.53$\pm$ 0.30 &   8.83$\pm$ 2.86\\
 0.30647 &   8349& \textit{b} & UVOT & $>$21.83 & $<$  6.76\\
 1.14624 & 127458& \textit{b} & UVOT & $>$22.19 & $<$  4.85\\
 2.44615 &  74224& \textit{b} & UVOT & $>$21.00 & $<$ 14.47\\
 4.07834 & 138632& \textit{b} & UVOT & $>$21.33 & $<$ 10.64\\
\hline        
 0.01789 &    318& \textit{v} & UVOT & $>$19.35 & $<$ 66.30\\
 0.07278 &   1431& \textit{v} & UVOT & $>$20.36 & $<$ 26.12\\
 0.18317 &   6538& \textit{v} & UVOT & $>$21.04 & $<$ 13.96\\
 0.31290 &   5819& \textit{v} & UVOT & $>$20.69 & $<$ 19.20\\
 1.11739		& 21$\times$180		& \textit{V}		& 1.23mCAHA				& 22.47$\pm$0.19	& 3.73$\pm$0.65	\\
 1.18728 & 121098& \textit{v} & UVOT & $>$21.08 & $<$ 13.39\\
 2.45087 &  74279& \textit{v} & UVOT & $>$20.50 & $<$ 22.93 \\
 4.08128 & 138537& \textit{v} & UVOT & $>$20.92 & $<$ 15.56 \\ 
 \hline
39.11207		& 6$\times$180		& \textit{g$^\prime$}		& OSIRIS/10.4mGTC	& $>$ 26.3	& $<$ 0.11	\\
39.49403		& 5$\times$180		& \textit{g$^\prime$}		& GMOS/8mGemini			& 26.80$\pm$0.35	& 0.07$\pm$0.03	\\
$\sim$180			&42$\times$200		&\textit{g$^\prime$}		&OSIRIS/10.4mGTC		& 27.21$\pm$0.27	&0.047$\pm$0.010	\\
 \hline
  1.04545		& 19$\times$180		& \textit{R}		& 1.23mCAHA				& 22.61$\pm$0.16	& 3.28$\pm$0.48	\\
 \hline
  0.29887		& 3$\times$300	 	& \textit{r$^\prime$}		& CQUEAN/2.1mMcD			& 22.43$\pm$0.14	& 3.87$\pm$0.50	\\
  2.08833		& 1$\times$30			& \textit{r$^\prime$}		& OSIRIS/10.4mGTC	& 23.39$\pm$0.12	& 1.60$\pm$0.18	\\
21.15017		& 10$\times$60		& \textit{r$^\prime$}		& OSIRIS/10.4mGTC			& 24.21$\pm$0.14	& 0.75$\pm$0.10	\\
28.49818		& 5$\times$180		& \textit{r$^\prime$}		& GMOS/8mGemini		& 24.81$\pm$0.13	& 0.43$\pm$0.05	\\
39.10159		& 4$\times$120		& \textit{r$^\prime$}		& OSIRIS/10.4mGTC	& 24.77$\pm$0.13	& 0.45$\pm$0.05	\\
39.47981		& 5$\times$180		& \textit{r$^\prime$}		& GMOS/8mGemini			& 25.24$\pm$0.15	& 0.29$\pm$0.04	\\
44.08258		& 4$\times$180		& \textit{r$^\prime$}		& OSIRIS/10.4mGTC	& $>$ 24.7	& $<$ 0.48	\\
$\sim$180			& 32$\times$200		&\textit{r$^\prime$}		& OSIRIS/10.4mGTC	& 26.90$\pm$0.14	&  0.063$\pm$0.008	\\
  \hline
  1.17359		& 17$\times$180		& \textit{I}		& 1.23mCAHA			& 22.18$\pm$0.35	& 4.88$\pm$1.57	\\
61.96267		& 20$\times$120		& \textit{I}		& SCORPIO/6mBTA				& 25.17$\pm$0.35	& 0.31$\pm$0.10	\\
   \hline
  0.29516		& 3$\times$300		& \textit{i$^\prime$}		& CQUEAN/2.1mMcD			& 22.72$\pm$0.18	& 2.96$\pm$0.49	\\
10.09449		& 9$\times$900		& \textit{i$^\prime$}		& RAT/2.0mLT				& 24.01$\pm$0.13	& 0.90$\pm$0.11	\\
39.12164		& 5$\times$60			& \textit{i$^\prime$}		& OSIRIS/10.4mGTC	& 24.36$\pm$0.17	& 0.65$\pm$0.11	\\
39.46336		& 5$\times$180		& \textit{i$^\prime$}		& GMOS/8mGemini			& 24.61$\pm$0.09	& 0.52$\pm$0.04	\\
  \hline
  0.30384		& 3$\times$300		& \textit{z$^\prime$}		& CQUEAN/2.1mMcD			& 22.65$\pm$0.34	& 3.16$\pm$1.00	\\
39.09432		& 6$\times$60			& \textit{z$^\prime$}		& OSIRIS/10.4mGTC	& 24.73$\pm$0.42	& 0.47$\pm$0.18	\\
39.44619		& 7$\times$180		& \textit{z$^\prime$}		& GMOS/8mGemini			& 24.77$\pm$0.25	& 0.45$\pm$0.10	\\
\hline 
  0.30745		& 3$\times$300		& \textit{Y}		& CQUEAN/2.1mMcD			& $>$ 22.5	& $<$ 3.63	\\
\hline
37.45092		& 32$\times$60		& \textit{J}		& NIRI/8mGemini		& $>$ 23.4	& $<$ 1.58	\\
\hline
28.46873		& 44$\times$60		& \textit{K$_S$}	& NIRI/8mGemini		& 24.48$\pm$0.35	& 0.59$\pm$0.19	\\
\hline                                   
\end{tabular}
\end{center}
}\end{table}

\subsection{McD 2.1m}
The CQUEAN instrument (Camera for QUasars in the EArly uNiverse; Park et al. 2011, in preparation) on the 2.1m Otto-Struve telescope at McDonald Observatory, Texas, USA, observed the optical counterpart starting at 01:16:23 UT, on Dec. 26, 2010 (6.64 h after the burst). Three exposures of 300 s each were taken in $r^\prime$, $i^\prime$, $z^\prime$, and $Y$ bands under photometric conditions. The data were reduced with standard procedures of dark and flat-field corrections. The afterglow is detected in the $r^\prime$, $i^\prime$, and $z^\prime$-band images, the $Y$ band gives an upper limit only.

\subsection{CAHA 1.23m}
The 1.23m telescope is located at the German-Spanish observatory of Calar Alto (CAHA) in Almer\'{\i}a, Spain and is equipped with an optical imaging camera. The optical counterpart was detected in the $VRI$ bands $1.04-1.11$ days after the GRB trigger. The 1.23m was also used to calibrate the object field in $BVRI$ bands by observing the Landolt fields RU149D and SA98 on Dec. 26 and 27, 2010, under photometric conditions.

\subsection{LT 2.0m}
The Liverpool telescope is a 2.0m fully robotic telescope located at the observatory  of Roque de los Muchachos on La Palma.  Observations were carried out with the imaging camera RATCAM. The optical counterpart was detected in one epoch in $i^\prime$-band at 10.09 d after the GRB.

\subsection{OSIRIS/GTC}

We  acquired  imaging data using   OSIRIS  at the  Gran Telescopio  de Canarias (GTC), a 10.4m telescope located at  the observatory of Roque de los Muchachos  on La Palma, Canary Islands,  Spain. 

The  observations started in  $r^\prime$  band $\sim$2 days after  the burst, exposing for 30 s.  A second $r^\prime$ observation was carried out $\sim 21$ days after the gamma-ray event, where 5 exposures of 180 s were obtained. We furthermore obtained a   late-time   SED at 39  days  in $g^\prime$, $r^\prime$,  $i^\prime$  and $z^\prime$  bands and  a last
image at $\sim$44 days in the $r^\prime$ band.

The data of our two late epochs (at  $\sim$39 and $\sim$44 days) were obtained at a considerable airmass (1.73 -- 2.14) since the object was setting  quickly after evening   twilight. The data at $\sim$21  and $\sim$44 days were acquired with the Moon at $\sim$54 and $\sim$36 degrees, respectively,  with an illumination of 83\% and  21\%, respectively.  The SED at 39 days was constructed in dark time. The observing conditions  were good  in all four GTC epochs.

\subsection{Gemini-North: NIRI and GMOS-N}
Late-time imaging of the optical counterpart of GRB\,101225A was conducted with the Gemini-North observatory on Mauna Kea/Big Island, Hawaii, on several occasions. On the night of Jan. 23, 2011  we observed the field with the Near InfraRed Imager (NIRI) in the $K^\prime$ filter for $44\times60$ s exposures ($2\times30$ s co-adds) before switching to the Gemini Multi-Object Spectrograph (GMOS-N) for $5\times180$ s exposures each in the $r^\prime$ filter.  On the night of Feb. 1, 2011 we re-observed the field with NIRI in the $J$ band for $32\times60$ s exposures (co-added), and finally on the night of Feb. 3, 2011 we imaged the field in all four GMOS broad-band filters ($g^\prime r^\prime i^\prime z^\prime$). Since the source was setting, all exposures were taken at moderate to high airmass ($1.5-2.5$), although under relatively good seeing conditions.

\subsection{BTA 6m }
A final late image was obtained using SCORPIO on the 6.0m BTA telescope, located at the Special Astrophysical Observatory, in Russia. The observation consisted of $20\times120$ s exposures using an $I$ filter obtained on Feb. 25, 2011 (60 days post burst) under good weather conditions and a seeing of $1.3-2.0$ arcsec.

\subsection{Photometry of ground-based data}
The photometry of \textit{V}, \textit{R} and \textit{I} data was done in a consistent way using a set of 15 comparison stars in the field of GRB\,101225A calibrated with the Landolt fields taken on Dec. 26 and 27, 2011 by the 1.23m CAHA. For \textit{r$^\prime$}, \textit{i$^\prime$} and \textit{z$^\prime$} photometric calibrations was done with observations form the 2.1m Otto-Struve telescope at McDonald Observatory on Dec. 26, 2010, using the standard star Feige 34. Finally, \textit{g$^\prime$} photometry was derived from the other  reference magnitudes using numerical transformations \cite{Jester2005}. The magnitudes of the comparison stars in the different filters used for the optical observations are listed in Tab.~\ref{table:compstars}.
  
We performed aperture photometry using PHOT within IRAF taking an aperture radius equal to the Full Width at Half Maximum (FWHM) of the stellar point sources. In cases where the contamination by neighboring sources was not negligible, we did PSF photometry within IRAF. In Tab.~\ref{table:log} we list the final photometry for all UV, optical and IR data.

     \begin{figure}[ht!]
   \centering
   \includegraphics[width=11cm]{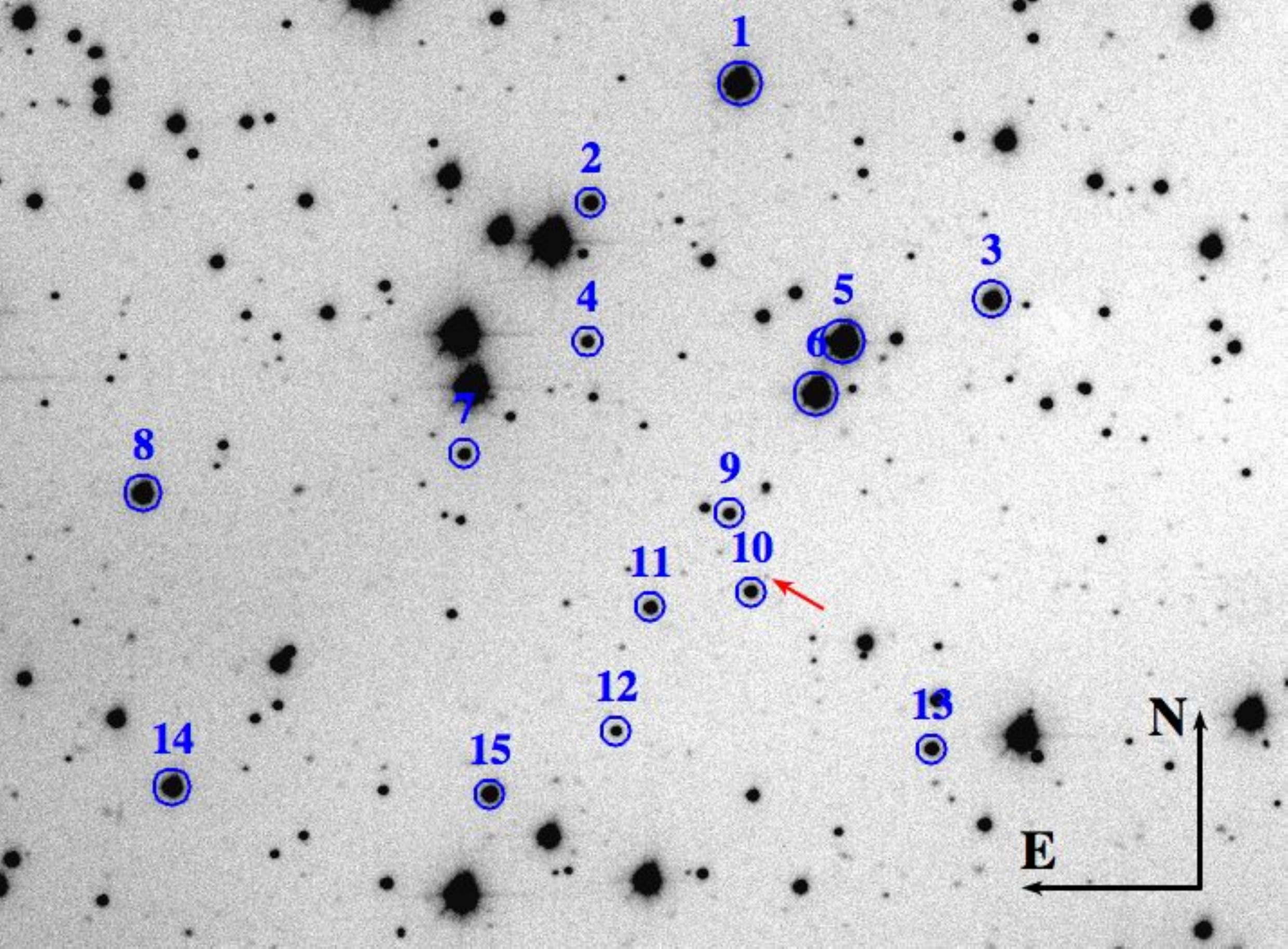}
      \caption{{\bf Field with the secondary standards used for the photometric calibration.} The position of the optical counterpart of GRB\,101225A is indicated with an arrow. The photometric magnitudes of each of the reference stars are given in Table~\ref{table:compstars}. The field of view is 6 arcmin $\times$ 4 arcmin.
              }
         \label{Fig:fc}
   \end{figure}

\begin{table*}[ht!]
\caption{Magnitudes of calibration stars used for the optical photometry. A finding chart indicating the position of each reference star is given in Fig.~\ref{Fig:fc}.  \textit{V}, \textit{R} and \textit{I} are given in Vega system, the other bands are in AB.  \label{table:compstars}}
{\scriptsize
\begin{center}
\begin{tabular}{llllllllll} \hline
ID  &R.A. (J2000)&Dec. (J2000)& \textit{V}                    &           \textit{R}           &         \textit{I}                &       \textit{g$^\prime$}          &      \textit{r$^\prime$}          &       \textit{i$^\prime$}          &        \textit{z$^\prime$}        \\ \hline
1 & 00:00:48.13&+44:38:18.9 & 15.282 $\pm$ 0.009&14.916 $\pm$ 0.049 &14.553 $\pm$ 0.052 & ------	&    ------        &     -----        &       -----      \\ 
2 & 00:00:52.02&+44:37:46.6 & 18.065 $\pm$ 0.097&17.268 $\pm$ 0.055 &16.288 $\pm$ 0.110 & 18.62$\pm$0.23 & 17.679$\pm$0.011 & 17.121$\pm$0.010 & 16.870$\pm$0.010 \\
3 & 00:00:41.58&+44:37:18.0 & 16.794 $\pm$ 0.050&16.421 $\pm$ 0.051 &15.972 $\pm$ 0.062 & 17.14$\pm$0.12 &    ------        &     -----        &       -----      \\
4 & 00:00:52.13&+44:37:07.9 & 18.789 $\pm$ 0.079&18.335 $\pm$ 0.133 &18.051 $\pm$ 0.162 & 19.26$\pm$0.19 & 18.464$\pm$0.012 & 18.325$\pm$0.011 & 18.263$\pm$0.013 \\
5 & 00:00:45.46&+44:37:06.6 & 14.995 $\pm$ 0.008&14.633 $\pm$ 0.049 &14.235 $\pm$ 0.051 & ------			&    ------        &     -----        &       -----      \\
6 & 00:00:46.18&+44:36:52.4 &15.174 $\pm$ 0.015&14.838 $\pm$ 0.049 &14.456 $\pm$ 0.051 & ------	&    ------        &     -----        &       -----      \\
7 & 00:00:55.37&+44:36:36.8 &19.165 $\pm$ 0.180&18.268 $\pm$ 0.090 &17.356 $\pm$ 0.188 & ------			& 18.569$\pm$0.012 & 17.905$\pm$0.011 & 17.597$\pm$0.011 \\
8 & 00:01:03.76&+44:36:26.8 & 16.594 $\pm$ 0.012&16.161 $\pm$ 0.049 &15.691 $\pm$ 0.090 &	------			&    ------        &     -----        &       -----      \\ 
9 &00:00:48.48&+44:36:19.3 &18.796 $\pm$ 0.059&17.924 $\pm$ 0.069 &16.694 $\pm$ 0.092 & 19.08$\pm$0.14	& 18.602$\pm$0.012 & 17.819$\pm$0.011 & 17.516$\pm$0.011 \\
10&00:00:47.98&+44:35:57.8 & 18.682 $\pm$ 0.132&17.901 $\pm$ 0.107 &17.059 $\pm$ 0.182 & 19.41$\pm$0.31& 18.174$\pm$0.011 & 17.645$\pm$0.010 & 17.427$\pm$0.012 \\
11&00:00:50.58&+44:35:43.5 & 18.505 $\pm$ 0.218&17.877 $\pm$ 0.060 &17.270 $\pm$ 0.163 & ------			& 18.088$\pm$0.011 & 17.715$\pm$0.010 & 17.537$\pm$0.012 \\
12 &00:00:51.59&+44:35:19.1&19.253 $\pm$ 0.130&18.893 $\pm$ 0.131 &18.460 $\pm$ 0.273 & 19.54$\pm$0.31& 19.059$\pm$0.014 & 18.910$\pm$0.013 & 18.852$\pm$0.020 \\
13 &00:00:43.29&+44:35:13.1&18.050 $\pm$ 0.066&17.749 $\pm$ 0.085 &17.318 $\pm$ 0.108 & 18.34$\pm$0.16 & 17.853$\pm$0.011 & 17.729$\pm$0.010 & 17.671$\pm$0.012 \\
14 &00:01:03.09&+44:35:04.8&16.828 $\pm$ 0.015&16.321 $\pm$ 0.052 &15.826 $\pm$ 0.087 & ------ 	&    ------        &     -----        &       -----      \\
15 &00:00:54.83&+44:35:01.9&17.516 $\pm$ 0.032&17.024 $\pm$ 0.066 &16.321 $\pm$ 0.082 & 18.04$\pm$0.08 & 17.148$\pm$0.010 & 16.960$\pm$0.010 & 16.853$\pm$0.011 \\

\hline
\end{tabular}
\end{center}
}
\end{table*}

\subsection{Preimaging}

\begin{figure}[ht!]
   \centering
   \includegraphics[width=8cm]{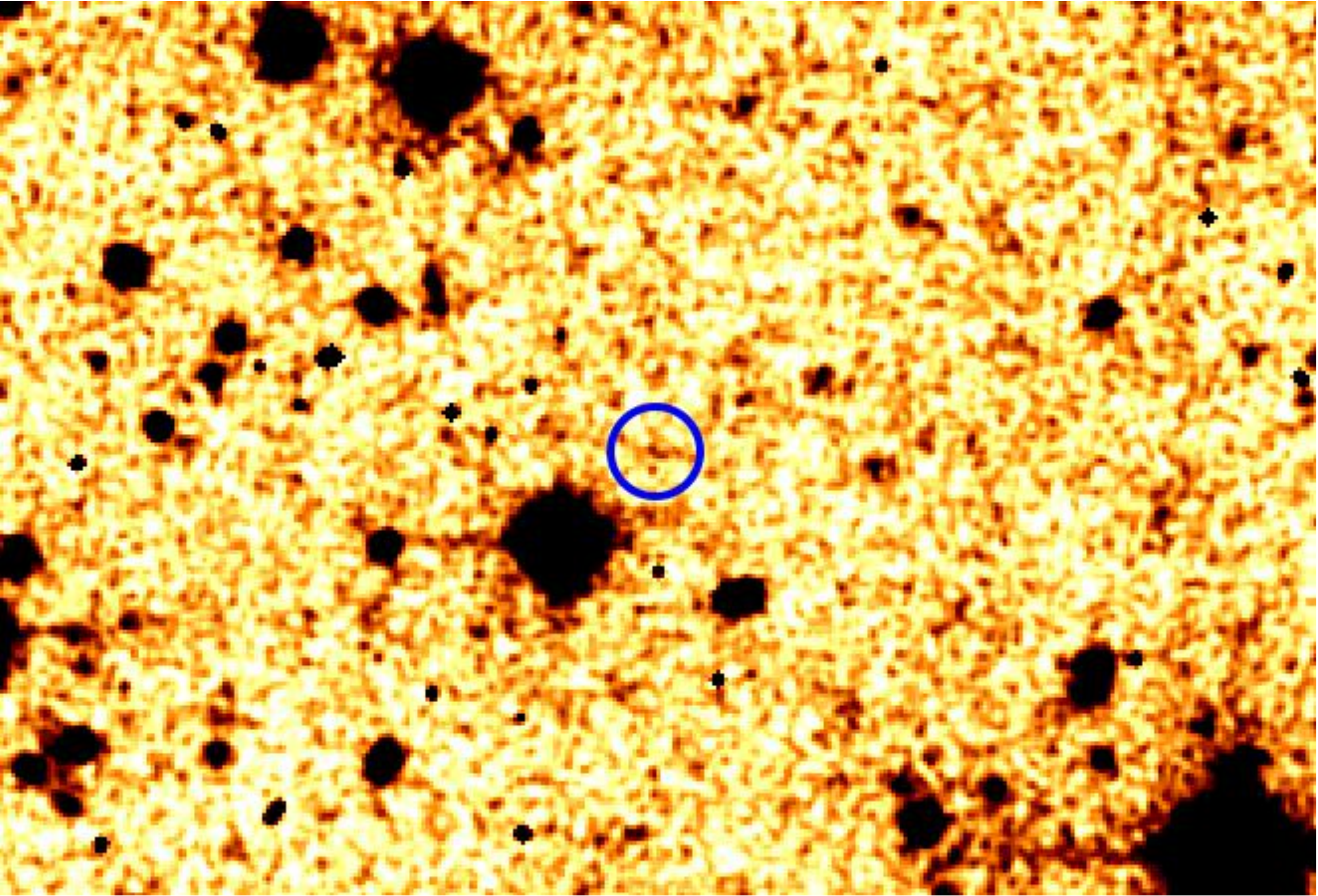}
      \caption{{\bf Pre-imaging exposure in $g^\prime$ band obtained with the 3.5m CFHT.} The field of view is 60 arcsec $\times$ 40 arcsec, North is to the top and East to the left of the image. The circle indicates the position of the optical counterpart of GRB\,101225A, where we see a 2$\sigma$ detection of what could be the host galaxy.
              }
         \label{Fig:preima}
   \end{figure}

Preimaging of the field was available from the archive of the 3.5m Canada-France-Hawaii Telescope (CFHT), obtained with the MegaPrime/MegaCam for the Pan-Andromeda Archaeological Survey (PAndAS, \cite{Richardson11}). We combined $3\times500$ s exposures obtained under very good conditions in $g^\prime$ and $i^\prime$ bands and derive $3\sigma$ limiting magnitudes for these exposures of $i^\prime>25.5$ and $g^\prime>26.9$. At the position of GRB\,101225A, we tentatively detect an object with $g^\prime=27.2\pm0.5$ (2$\sigma$, see Fig. \ref{Fig:preima}) which is consistent with our late host detection (see Sect. \ref{host}).

\subsection{Host detection}\label{host}
At three epochs, June 9, June 28 and 29, we obtained deep imaging of the field of GRB 101225A in g$^\prime$ and r$^\prime$ with OSIRIS at the 10.4m GTC under good conditions and seeing around 1.0 arcsec. The total exposure time is 8400\,s in $g^\prime$ and 6400\,s in $r^\prime$ band. We detect an unextended object at the position of the GRB in both bands at $g^\prime=27.21\pm0.27$ ($\sim$3$\sigma$) and $r^\prime=26.90\pm0.14$ ($\sim$7$\sigma$). These values are clearly above the extrapolation of the SN light curve to 180 days and we therefore propose this object to be the host galaxy of GRB 101225A. At a redshift of $z=0.33$, the absolute luminosity is M$_\mathrm{abs,g}=-13.7$ mag, $\sim$\,2\,mag fainter than the faintest GRB host detected so far (XRF 060218 [12]) and one of the faintest galaxies ever detected at that redshift. The $g^\prime-r^\prime$ color of the host is consistent with what we would expect for a late-type star-forming galaxy, in line with what is seen for other long GRBs, although we note that the large errors allow other galaxy types as well. The blue color would also argue against the source being a late detection of a cool component from the event itself (see [3]).


\begin{figure}[ht!]
   \centering
   \includegraphics[width=13cm]{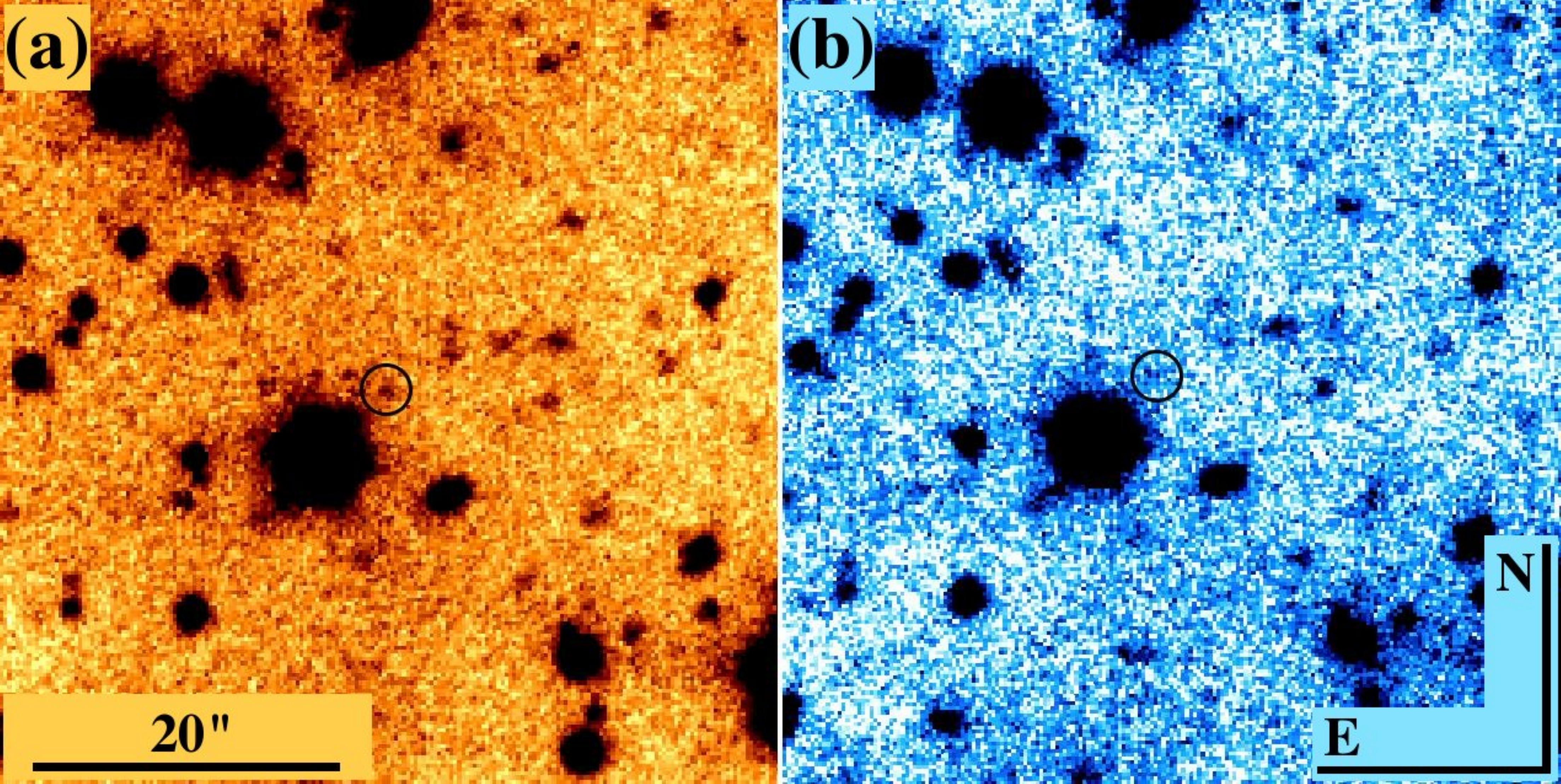}
      \caption{ {\bf Imaging of the host using OSIRIS/GTC  6 months after the GRB}. Panel a shows the stacked image in $r^\prime$, panel b the same in the $g^\prime$ band. The field of view is 50 arcsec $\times$ 65 arcsec. The circle indicates the position of the host galaxy.}
         \label{Fig:host}
   \end{figure}

\section{Optical spectroscopy}
\label{sec:spec}
We obtained a spectrum of the optical counterpart 51 h after the event using OSIRIS on the 10.4m GTC telescope on La Palma (Spain). Two spectra of 1800 s exposure time each were taken with grism 300B (R$=$325, wavelength range: $3500-7000$ {\AA}) under moderate to high airmass (1.26 and 2.05, respectively). The spectra were reduced and combined with standard tasks in IRAF and flux-calibrated with the spectrophotometric standard G191-2B2 observed in the same night. The continuum is clearly detected, but the spectrum shows no obvious absorption or emission lines.

We performed a search for any possible redshift solution in the range z$=$0.1 -- 0.6. For this, we stacked the normalized spectrum at the position of the strongest emission lines [OII] $\lambda\lambda$ 3727, 3729 \AA{}, [OIII] $\lambda\lambda$ 4959, 5007 \AA{}, H$\beta$ and H$\alpha$ at redshifts between 0.1 and 0.6 in steps of 0.005. We find no indication of any emission feature in the stacked spectrum at any redshift in this interval.

\begin{figure}[h]
   \centering
   \includegraphics[width=\columnwidth]{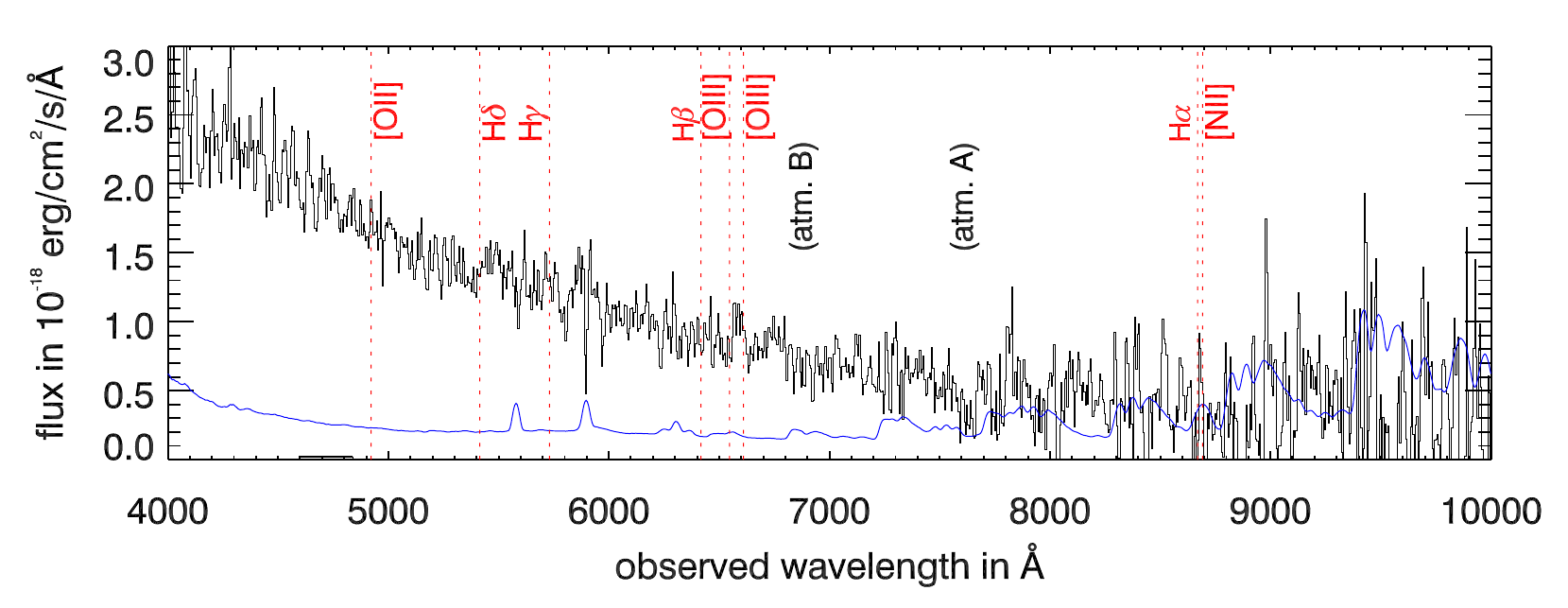}
      \caption{{\bf Flux-calibrated spectrum of the optical transient from GTC/OSIRIS 2.1 days after the GRB}. The error spectrum is plotted in blue. The red lines indicate the position of normally strong emission lines from the interstellar medium at a redshift of $z=0.33$, none of the lines is detected in our spectrum.
              }
         \label{Fig:specGTC}
   \end{figure}

The limits on the detection of H$\alpha$ [OIII] and [OII] emission from the host galaxy are $<$~5$\times$10$^{-18}$ erg\,cm$^{-2}$\,s$^{-1}$, $<$~2.3$\times$10$^{-18}$\,erg\,cm$^{-2}$\,s$^{-1}$ and $<$~3$\times$10$^{-18}$\,erg\,cm$^{-2}$\,s$^{-1}$ (3$\sigma$) respectively. We can also put a limit on the detection of H$\alpha$ at $z=0$ of $<$~2$\times$10$^{-18}$\,erg\,cm$^{-2}$\,s$^{-1}$. The flux-calibrated spectrum with the position of typical emission lines from the host shifted to a redshift of $z=0.33$ is shown in Fig. \ref{Fig:specGTC}.

On the night of 2011 Feb. 04 we observed the optical counterpart with the Low-Resolution Imaging Spectrometer \cite{Oke95} on the Keck I telescope during local twilight.  Two undithered observations of 600 seconds each were acquired using the 1.0 arcsec slit and the D500 dichroic at a position angle of 86.5 degrees.  On the red arm we used the 600/7500 grating and binned the CCD along the spatial direction (2x1 binning); on the blue arm we used the 600/4000 grism and binned the data along both spatial and spectral axes (2x2 binning).  Due to twilight there is no evidence of a trace in the blue spectrum (and no source is detected in GMOS $g^\prime$-band imaging from the previous night). A faint continuum trace at the expected position of the transient is identified on the red side in the second (less twilight-affected) exposure from 7160 to 8000 \AA{} with no absorption or emission features visible. 

\section{Modeling the UV to NIR spectral energy distribution}

The early evolution of the UV/optical/IR (UVOIR) counterpart is very unusual for a GRB afterglow. Instead of a power-law spectrum with a negative spectral slope, it had a very blue counterpart, following what seemed to be a power-law with a positive spectral slope \cite{Cenko10}. Furthermore, the counterpart stayed bright during the first days and then decayed \cite{Xu10} with a strong color change, transforming into a very red counterpart two weeks after the trigger \cite{Tanvir11}. We interpret this early evolution as being produced by the expansion and cooling of a blackbody (BB), as shown in Section \ref{sec:sed}.

The simple BB evolution is not valid any more for the emission beyond $\sim20$ days after the trigger. At that time we observe a flattening of the light curve, while the very red color is preserved. This late evolution can be well-described with the presence of a supernova component, as described in Section \ref{sec:sn}.

\subsection{Early time evolution}
\label{sec:sed}

For the modeling of the UVOIR spectral energy distribution (SED), we use the photometry presented in Tab. \ref{table:log} together with some of the data points extracted from the literature \cite{Cenko10, Xu10, Tanvir11, Xu496, Wiersema10, Xu519, Fynbo11}. All magnitudes are corrected for a Galactic extinction of $A_V=0.33$ mag and transformed from magnitudes to flux densities. With this data set we are able to derive a set of 12 SEDs ranging from 0.07 to 40 days after the trigger.

The early optical SEDs are well fitted by using an expanding and cooling blackbody of the following form (in frequency space):
\begin{equation}
F_{\nu} (Jy) =  10^{26}\left(\frac{R}{D}\right)^2 \frac{2\pi h \nu^3(1+z)^4}{c^2} \frac{1}{e^{h\nu/k_BT_{obs}}-1}
\end{equation}
Here, the factor $10^{26}$ is used to convert $W/m^2/Hz$ to Jy. $R$ is the radius of the emitting black body (which we assume to be spherical), $D$ is the luminosity distance to the object, $z$ the redshift and $T_{obs}$ is the observed blackbody temperature (the rest-frame temperature would be $T_{rest}=T_{obs}(1+z)$). The other physical constants are: $c$ is the speed of light, $h$ Planck's constant and $k_B$ Boltzmann's constant. For simplicity we assume a blackbody with an emissivity of 1. The blackbody succeeds in reproducing the data up to 10 days, without any intrinsic extinction or additional emission component, after which another component becomes dominant (see Section \ref{sec:sn}).

\begin{table}[ht!]
 \caption{Measured values for the blackbody evolution. Values in brackets are extrapolated from the BB evolution due to a limited amount of data points in those SEDs.}             
\label{table:bbfits}      
\centering                          
\begin{tabular}{c c c}        
\hline                 
Epoch      	& Observed temperature 			&  Normalisation constant	\\
(days)		&	(K)	    			&         $(1+z)^4\pi10^{26}\left(\frac{R}{D}\right)^2 $  		 \\
\hline                        

0.07  &   (43 000$\pm$8 000  & 1.7$\pm$1.5) \\
0.17  &   40 000$\pm$6 000  & 1.8$\pm$1.2  \\
0.3     &  35 000$\pm$3 950 &  2.2$\pm$0.8 \\
0.6    &   25 340$\pm$5 440 &  4.6$\pm$2.5 \\
1.1    &   20 900$\pm$1 770 &  5.2$\pm$1.2 \\
2.0    &   15 000$\pm$1 090  & 8.9$\pm$1.8 \\
3.0    &   14 260$\pm$1 760   & 8.00$\pm$3.4 \\
5.0    &   (11 300$\pm$2 000 &  10.0$\pm$4.0) \\
10.0  &   (6 000$\pm$2 000   &  14.0$\pm$4.0) \\ 
18.0  &   (5 000$\pm$1 000   &  18.0$\pm$10.0) \\ 
\hline
\end{tabular}
\end{table}

From the fits to the SED evolution and allowing for a second-order fit, we get the following evolution of the normalisation constant:
\begin{equation}
log\left(10^{26}\pi(1+z)^4\left(\frac{R}{D}\right)^2\right)=(0.70\pm0.04)+(0.46\pm0.03)log(t)-(0.01\pm0.05)log10(t)^2
\end{equation}
where $t$ is the time in days. The temperature evolution (in K) can be described by:
\begin{equation}
log\left(T_{obs}\right)=(4.342\pm0.017)-(0.395\pm0.016)log(t)-(0.11\pm0.02)log(t)^2
\end{equation}

\begin{figure}[!ht]
   \centering
   \includegraphics[width=11cm]{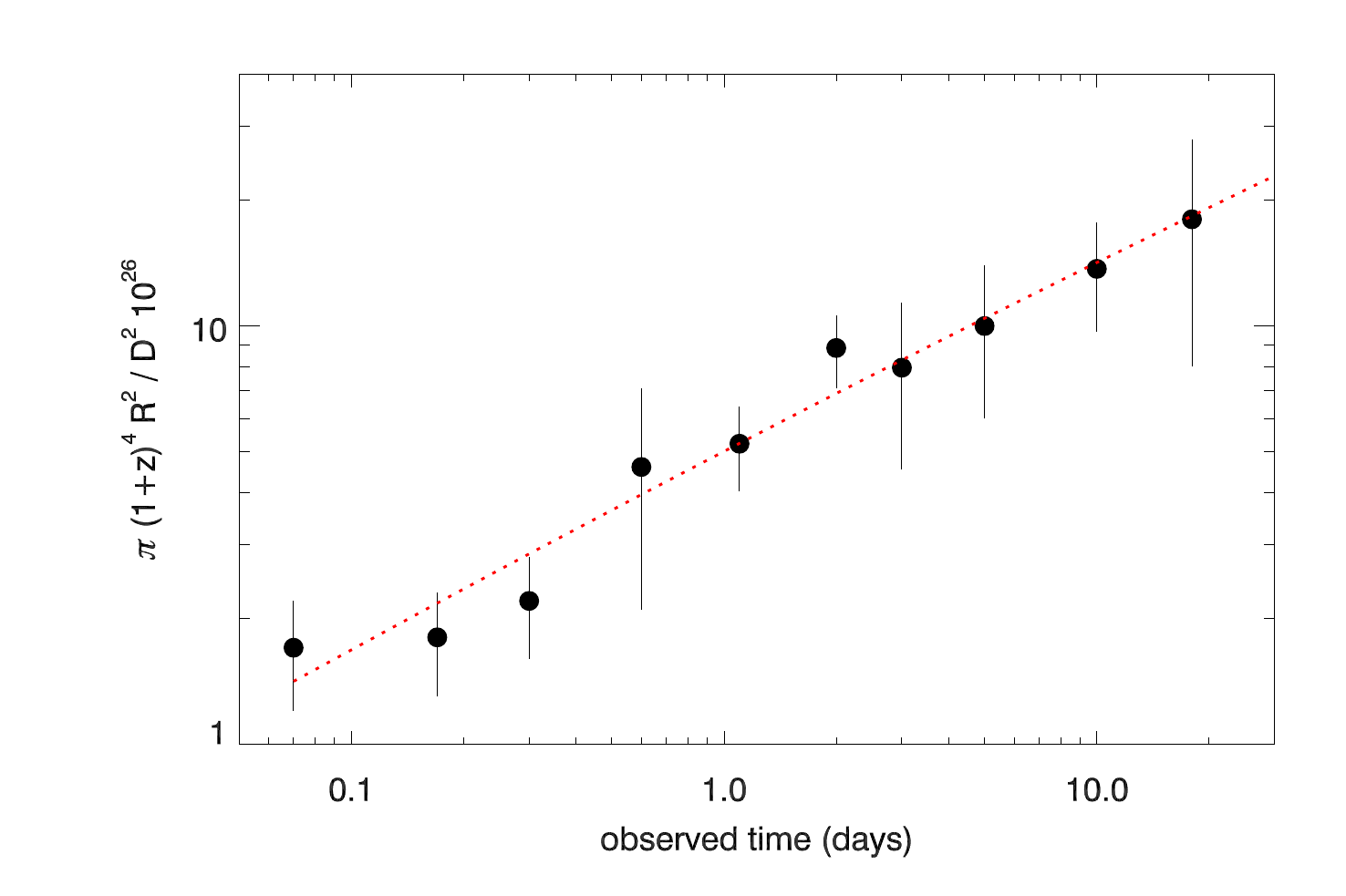}
    \includegraphics[width=11cm]{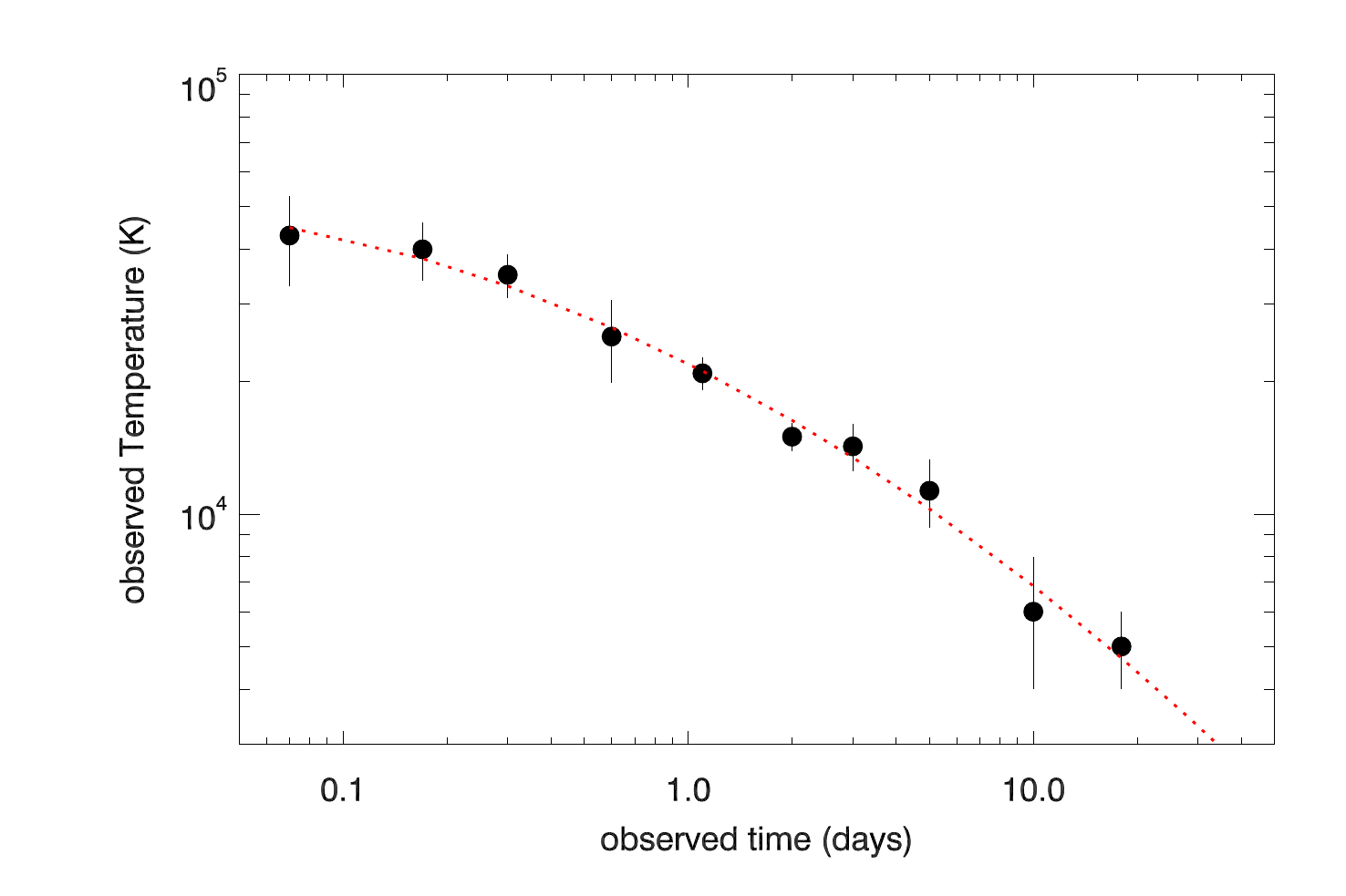}
      \caption{{\bf Temporal evolution of the normalization constant and temperature during the UVOIR BB phase}. The evolution of the normalization constant can be described by a linear evolution in log-log space while the temperature evolution requires a second order term.
              }
         \label{Fig:temp}
   \end{figure}

\begin{figure}[!ht]
   \centering
   \includegraphics[width=11cm]{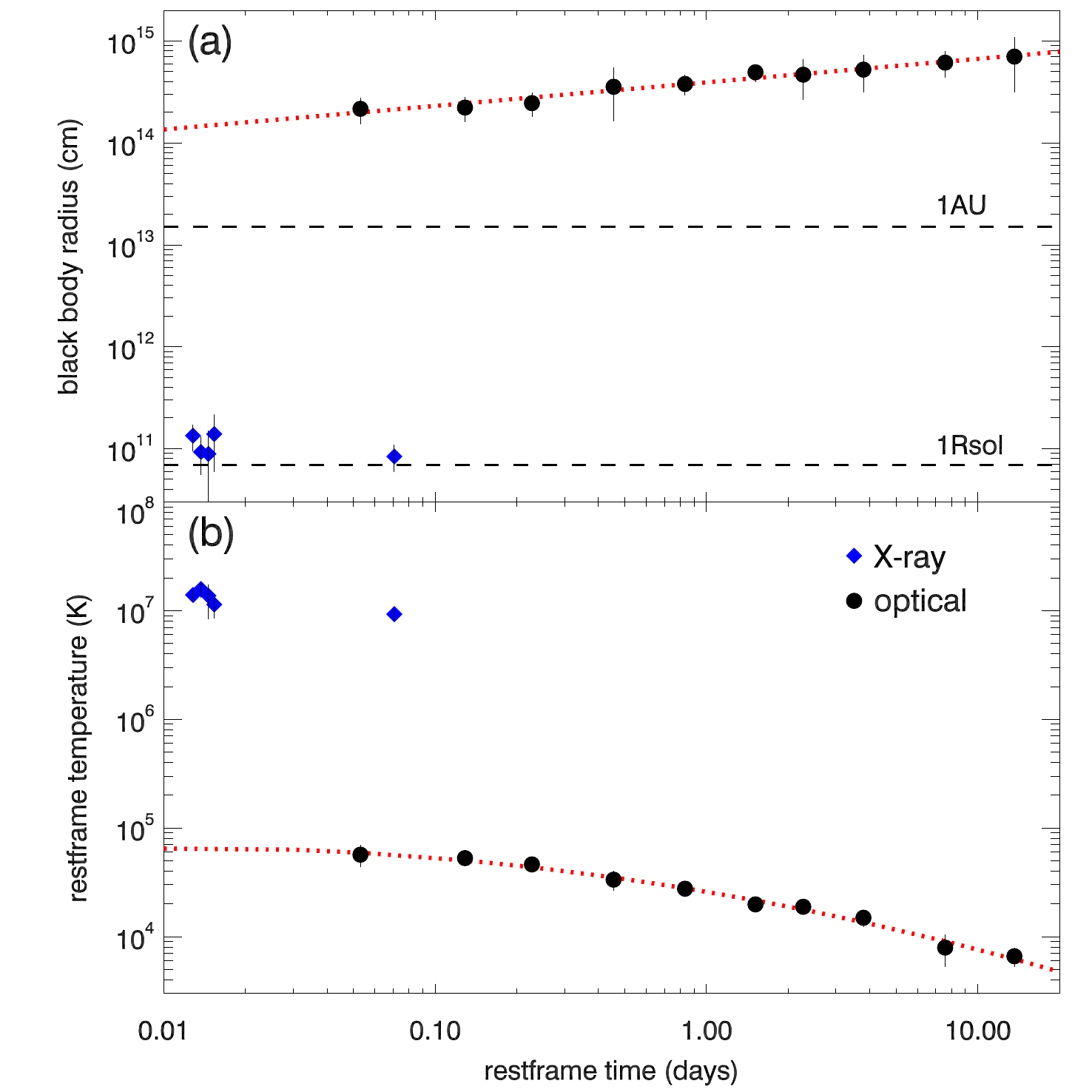}
      \caption{{\bf Temporal evolution of the physical radius and temperature of X-ray and UVOIR BB}. This figure is identical to \ref{Fig:temp} but showing the evolution of both BB components together.
              }
         \label{Fig:cte}
   \end{figure}

Figure \ref{Fig:temp} shows the temporal evolution of the normalisation constant and the temperature. The normalisation can be sufficiently described by a linear evolution in log-log space and therefore the second order term in eq. (2) can be neglected. For the temperature, we need an additional second order term to obtain a reasonable fit to the data. The temperature and normalization evolution fits very well to our theoretical model of the afterglow (see Sect. \ref{sect:model} and Fig.~\ref{Fig:Tevoltheory}). Once the redshift is known the normalisation constant can be transformed into physical values in the rest-frame of the object which is shown in Fig. \ref{Fig:cte}. For this we assume $z=0.33$ (see Section \ref{sec:sn}), or 1661.1 Mpc using a $\Lambda$CDM cosmology with $H_0=71$, $\Omega_M=0.27$ and $\Omega_\Lambda=0.73$. 
   
%

        \begin{figure}[!ht]
   \centering
   \includegraphics[width=11cm]{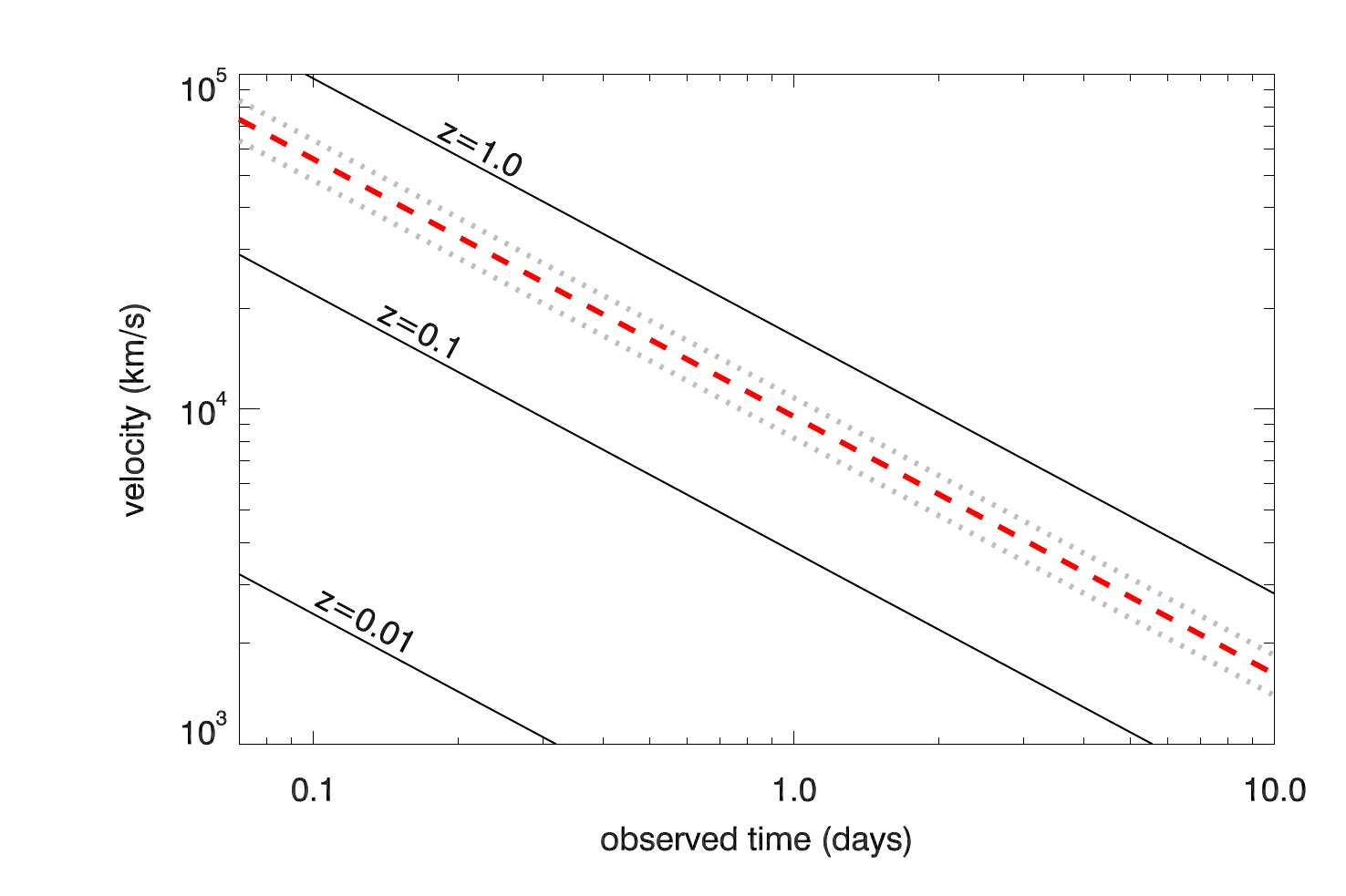}
      \caption{{\bf Evolution of the velocity of the UVOIR black body according to the result of our modeling.} The plot shows the evolution at the best-fit redshift of $z=0.33^{+0.07}_{-0.04}$ (red dashed line and grey dotted lines) as well as three other redshifts. At $z=0.33$, the velocities derived during the BB phase range between $0.25-0.005c$, typical of accelerated material during a supernova explosion. 
              }
         \label{Fig:vel}
   \end{figure}

\subsection{Late evolution and SN template fitting}
\label{sec:sn}

The late evolution of the light curve requires a component in addition to the evolving BB described in Section \ref{sec:sed}. A pure BB fit to the late data gives a bad fit with $\chi^2/d.o.f.=10.45/5$ and is therefore ruled out with a 94\% probability. We assume the late SED to have an additional component from a SN and fit the SED with several SN templates. These fits are also used to estimate the redshift of GRB\,101225A, which we could not obtain spectroscopically (Section \ref{sec:spec}).

\begin{figure}[ht!]
   \centering
   \includegraphics[width=12cm]{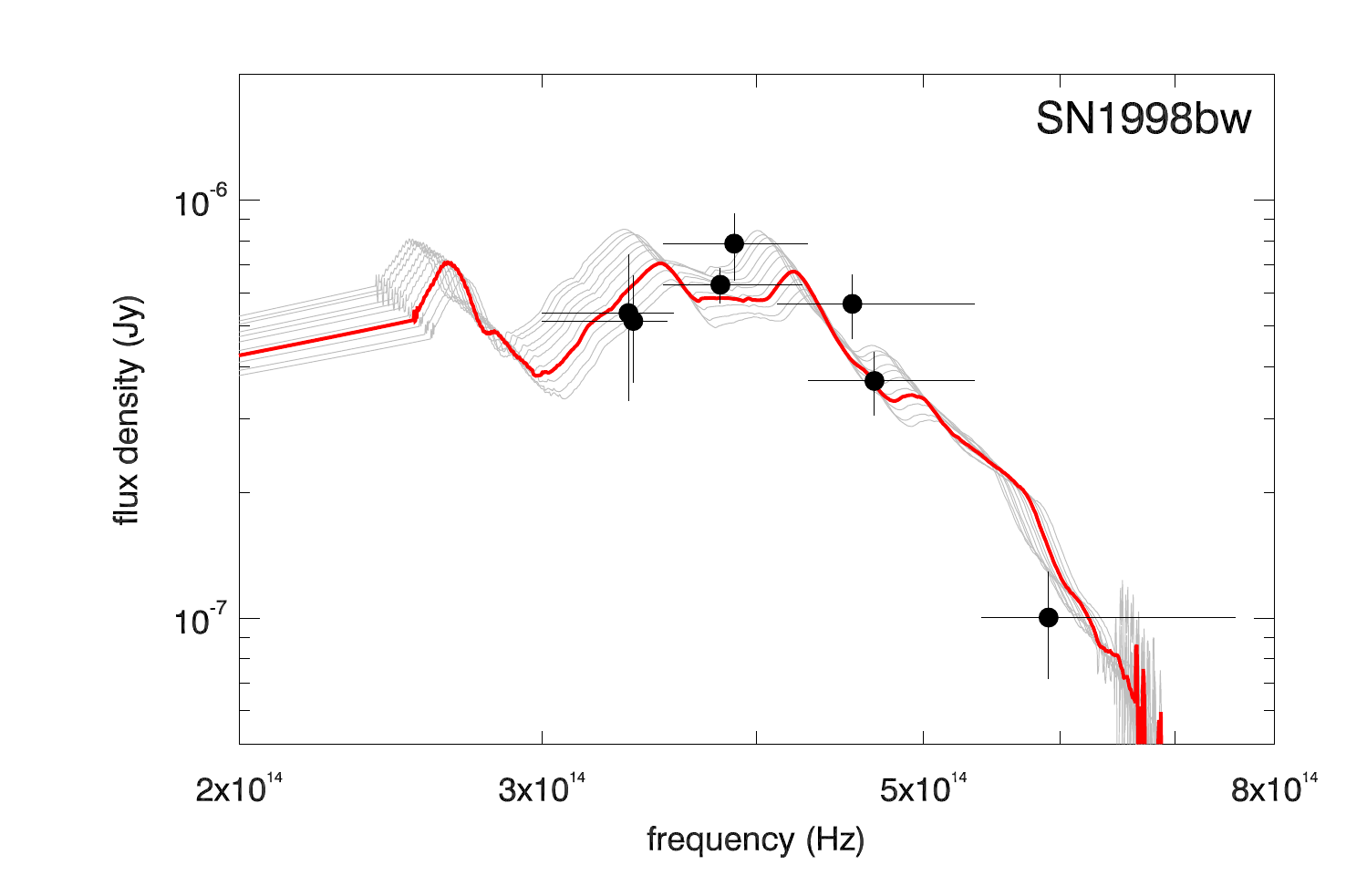}
      \caption{{\bf Fit of the day 40 SED to a SN 1998bw template.} The observations are in black and the best fit, with a redshift of $z=0.33$, in red. The gray lines represent the template at the different redshifts (steps of 0.01) within errors.
              }
         \label{Fig:98bw}
   \end{figure}

To determine the redshift of GRB\,101225A we use the SED at 40 days after the burst where the contamination from the BB is negligible and where we have detections in 7 different bands. Given the steep slope in the blue part of the spectrum, we convolve the response of each filter with the spectral shape of the templates.
This is particularly important for the $r^\prime$-band observations performed from GTC and Gemini at a very similar epoch, which show a significant difference in flux density. The filter of GTC reaches slightly redder wavelengths, and the difference in flux densities can be well  explained by a very steep slope due to a SN feature as shown in Fig. \ref{Fig:98bw}.

We obtain templates for different core-collapse supernovae from the literature \footnote{  http://supernova.lbl.gov/$\sim$nugent/nugent\_templates.html}. We exclude SN Ia from the analysis, as we do not expect high-energy emission such as detected for GRB 101225A for those events. The template for each SN prototype was interpolated to the time of the SED for a range of redshifts (see references in Table \ref{table:snfits}). In the particular case of SN 1998S the templates were created by combining ground-based and HST spectra.

In order to obtain the best-fit redshift, we also need to consider the time evolution and maximum of the SN light curve which is expressed as the stretching factor (a stretching factor of $s=1$ corresponds to a time evolution identical to the corresponding template SN). To this end, we iteratively fit the SED and the light curve starting with $s=1$ and a first fit to the SED in 7 bands (see Fig. \ref{Fig:98bw}). With the best-fit $z$, a SN light curve is derived and fitted to the real light curve which gives a new stretching factor. This process continues until the solutions for $s$ and $z$ converge. We also allow for a scaling of the flux of the SN during the fit. The best fits for $s$ and $z$ for each SN template are displayed in Table \ref{table:snfits}.

\begin{table*}[ht!]
 \caption{Fits of the SED with SN templates. A pure BB fit (not listed in this table) is rejected with a probability of 94\%. }             
\label{table:snfits}      
\centering                          
\begin{tabular}{l c c c c c c}        
\hline                 
SN Type         & SN template                   &  best fit $z$          & stretching factor               & $\chi^2/d.o.f.$               & rejection prob.& reference     \\
\hline
Ib/c                    & 1999ex                                & 0.31$^{+0.05}_{-0.10}$                & 1.1$\pm$0.15                  & 5.80/5             &    67\%         & {\small \cite{Stritzinger02}}\\
\textbf{Ic broad-lined}         & \textbf{1998bw }                      & \textbf{0.33$^{\mathbf{+0.07}}_{\mathbf{-0.04}}$}             & \textbf{1.25$\pm$0.15}                &\textbf{3.60/5}      &39\%                   & \textbf{{\small \cite{Galama98,Patat01}}}\\ 
II                      & 1998S                                 & 0.50$^{+0.07}_{-0.08}$                & 1.0   $\pm$0.15               & 44.5/5       &    100\%               & {\small \cite{Anupama01, Lentz01, Fassia01}} \\
IIN                     &  2001aj                                     & ---                                           & ---                   & $>$ 50.0/5                &         100\%       &  {\small \cite{Stoll11}}\\
IIL                     & 1985P                         & 0.43$^{+0.04}_{-0.05}$                & 1.3   $\pm$0.15               & 6.1/5           &        70\%       & {\small \cite{Gaskell92}}\\
IIP                     & 1999em                                & 0.41$^{+0.05}_{-0.03}$                & 1.0   $\pm$0.15               & 6.3/5     &         72\%            & {\small \cite{Leonard02}}\\
 \hline
\end{tabular}
\end{table*}

\begin{figure}[ht!]
   \centering
   \includegraphics[width=12cm]{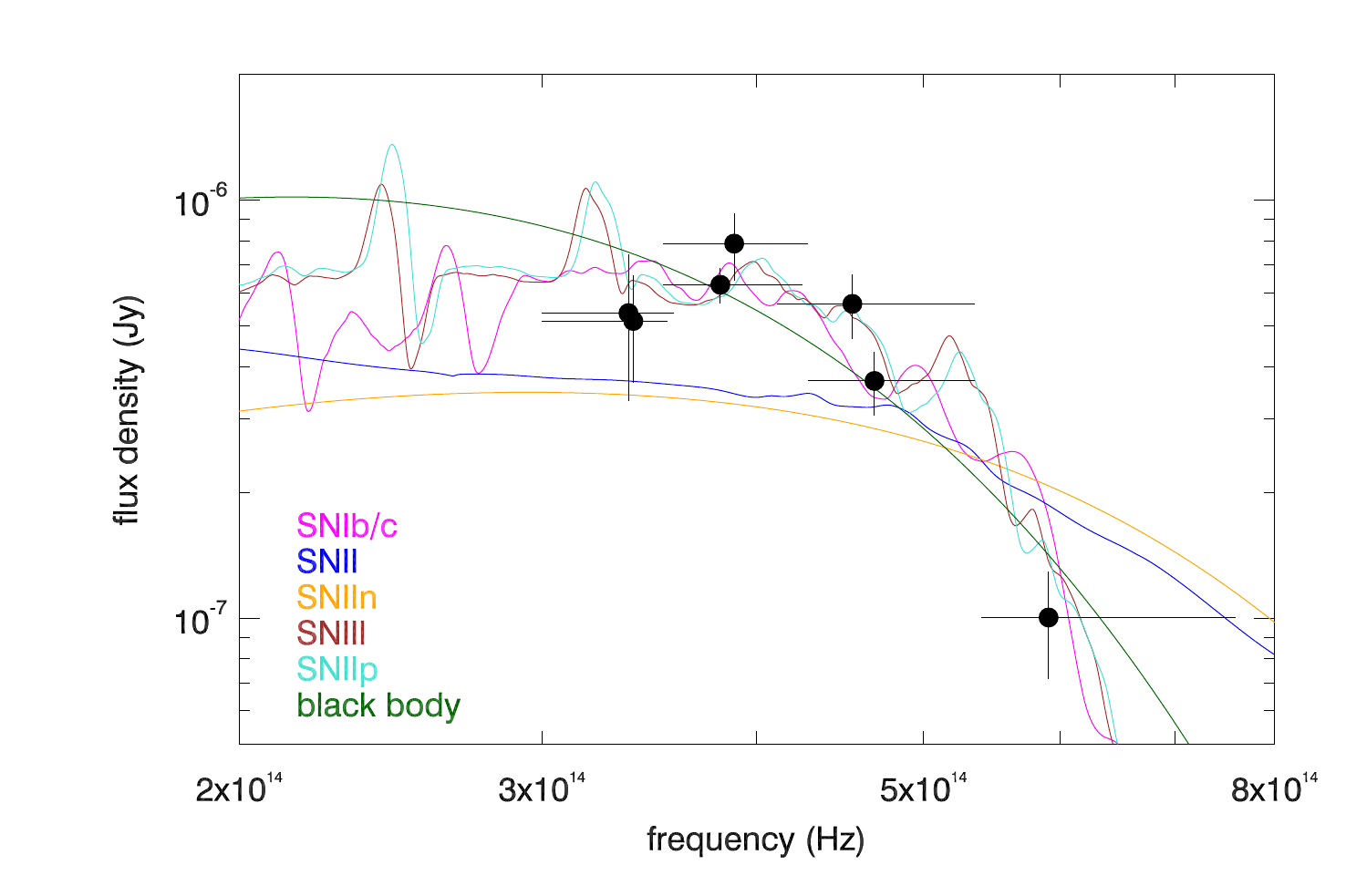}
      \caption{{\bf Fit of the SED at 40\,days to SN templates other than a SN 1998bw-like} (as listed in Tab. \ref{table:snfits}) The fit shown is the one with the best-fit stretching factor for each template. We also plot a simple BB with 3650 K, the temperature at which we obtain the best fit. 
              }
         \label{Fig:Snother}
   \end{figure}

The absolute best fit is obtained with a SN 1998bw template, a broad-lined Type Ic that is the canonical template for GRB-related supernovae. For this case we obtain a redshift of $z=0.33^{+0.07}_{-0.04}$ and $s=1.25\pm0.15$. The other core-collapse SNe we tested (excluding the Type II SN 1998S, which clearly does not fit our SED) give redshifts between $z=0.31$ and $z=0.50$. We therefore use $z=0.33$ as reference in this work. Figure \ref{Fig:98bw} shows the fitted SED with the template at $z=0.33^{+0.07}_{-0.04}$ and Figure \ref{Fig:Snother} the best fits for all other SN templates as well as a simple BB.

\subsection{Luminosity and stretching factor of the SN associated with GRB\,101225A in context of other GRB-SNe}   

We undertake a more general comparison to SNe associated with GRBs by following the formalism of \cite{Zeh04}. These authors used a SN 1998bw template light curve to fit late bumps in GRB afterglow light curves, modifying the template by increasing or decreasing the luminosity at peak (the parameter $k$, with $k=1$ implying a peak luminosity identical to that of SN 1998bw), and stretching or compressing the light curve in time while retaining the overall shape (the parameter $s$, again, $s=1$ implies the temporal evolution is identical to that of SN 1998bw in the same band). This procedure also included the creation of synthetic templates by interpolating between the SN 1998bw light curve in different filters, and taking into account the cosmological $K$-correction. Nearly all GRB-SNe were well-fit by the SN 1998bw light curve template. For GRB\,101225A we have $s=1.25\pm0.15$ and $k=0.08\pm0.03$ according to the designation of \cite{Zeh04} using the SN 1998bw light curve at $z=0.33$.

Ferrero et al. \cite{Ferrero06} analyzed SN 2006aj associated with XRF 060218, and placed it into the $k-s$ context. They employed the line-of-sight extinction values derived by \cite{Kann06} to derive intrinsic $k$ values. To place the SN associated with GRB\,101225A into the $k-s$ context, we fit the light curve analogous to \cite{Zeh04}, and use the sample of \cite{Ferrero06} as well as additional events as a comparison. The complete data are presented in Table \ref{table:ksgrbsne}.

GRB 990712 has been analysed again with additional data. We find no evidence for host extinction. For GRB 021211, a re-analysis of the afterglow SED finds no evidence for host extinction, the value from \cite{Ferrero06} thus remains unchanged but now counts as extinction-corrected. For GRB 040924, we use the $k$ and $s$ values from [9] and correct $k$ with the extinction found by \cite{Kann06}. For GRB 050525A, we use the uncorrected $k$ value from \cite{Ferrero06}, and correct it with the extinction found by [11]. For XRF 050824, we use the uncorrected $k$ value from \cite{Sollerman07}, and correct it with the extinction found by [11]. GRB 060729 is analysed in \cite{Kann11}. GRB 080319B is analysed in \cite{Bloom09}. GRB 090618 has been analysed for this work, using the data set of [18]. We were not able to derive a good SED for this afterglow, therefore the $k$ value has not been corrected.

\begin{figure}[ht!]
      \centering
   \includegraphics[width=9cm]{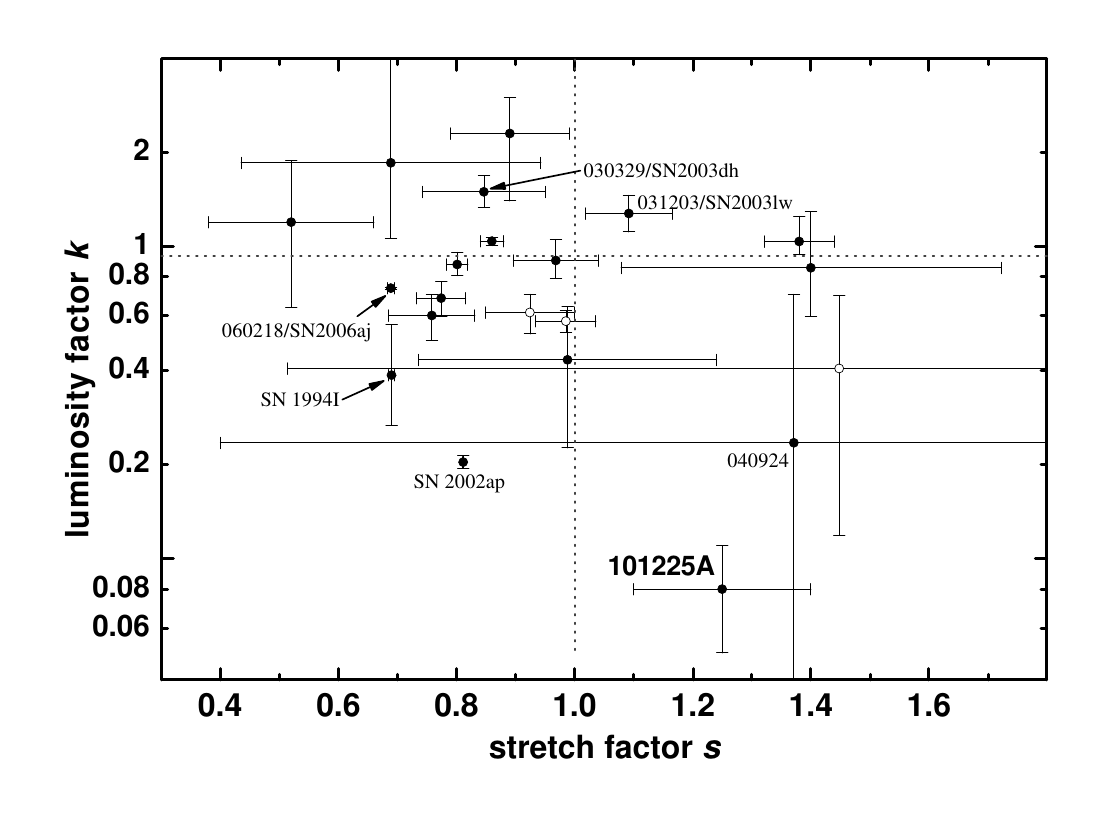}
      \caption{{\bf Luminosity factor $k$ and stretching factor $s$ of SNe associated with GRBs.} Filled symbols have been corrected for host-galaxy line-of-sight extinction, non-filled symbols have not. We label several well-studied nearby GRB-SNe, as well as two ``canonical'' Type Ic SNe, SN 1994I (Ic) and SN 2002ap (broad-lined Ic unassociated with a GRB). The GRB\,101225A SN is fainter than all these events.
              }
         \label{Fig:ks}
   \end{figure}

As can be seen in Fig. \ref{Fig:ks}, the SN associated with GRB\,101225A is significantly fainter than any other known GRB SN (with the SN associated with GRB 040924 being the most similar, but this event is only marginally detected). At the same time its temporal evolution is similar but slightly slower than most known GRB-SNe, though not by a large amount. It is also fainter than two well-studied Type Ic SNe, SN 1994I and SN 2002ap (see \cite{Ferrero06} for discussion), the latter being broad-lined, but not associated with a GRB.

\begin{table*}[ht!]
\caption{Luminosity Factor $k$ and Stretching Factor $s$ for GRB SNe}
\label{table:ksgrbsne}
\centering                          
\begin{tabular}{l c c l l }        
\hline                 
GRB	&	$k$					&	$s$			&	Comment	&	Reference	\\ \hline
970228	& $	0.41	\pm{	0.29			}$ & $	1.45	\pm	0.93	$ &	uncorrected	&	\cite{Ferrero06}	\\
990712	& $	0.60	\pm{	0.10			}$ & $	0.76	\pm	0.07	$ &		&	This Work	\\
000911	& $	0.85	^{+	0.44	}_{-	0.26	}$ & $	1.40	\pm	0.32	$ &		&	\cite{Ferrero06}		\\
010921	& $	1.85	^{+	2.82	}_{-	0.79	}$ & $	0.69	\pm	0.25	$ &		&	\cite{Ferrero06}		\\
011121	& $	0.88	^{+	0.08	}_{-	0.07	}$ & $	0.80	\pm	0.02	$ &		&	\cite{Ferrero06}		\\
020405	& $	0.90	^{+	0.15	}_{-	0.11	}$ & $	0.97	\pm	0.07	$ &		&	\cite{Ferrero06}		\\
020903	& $	0.62	\pm{	0.09			}$ & $	0.92	\pm	0.08	$ &	uncorrected	&	\cite{Ferrero06}		\\
021211	& $	0.43	\pm{	0.21			}$ & $	0.99	\pm	0.25	$ &		&	\cite{Ferrero06}	, This Work	\\
030329	& $	1.50	^{+	0.19	}_{-	0.16	}$ & $	0.85	\pm	0.10	$ &		&	\cite{Ferrero06}		\\
031203	& $	1.28	^{+	0.18	}_{-	0.16	}$ & $	1.09	\pm	0.07	$ &		&	\cite{Ferrero06}		\\
040924	& $	0.24	^{+	0.47	}_{-	0.23	}$ & $	1.37	\pm	0.97	$ &		&	[10], \cite{Kann06}	\\
041006	& $	1.04	^{+	0.22	}_{-	0.09	}$ & $	1.38	\pm	0.06	$ &		&	\cite{Ferrero06}		\\
050525A	& $	0.68	^{+	0.09	}_{-	0.08	}$ & $	0.77	\pm	0.04	$ &		&	\cite{Ferrero06}	, [11]	\\
050824	& $	1.20	^{+	0.69	}_{-	0.56	}$ & $	0.52	\pm	0.14	$ &		&	\cite{Sollerman07} , [11]	\\
060218	& $	0.74	\pm{	0.01			}$ & $	0.69	\pm	0.01	$ &		&	\cite{Ferrero06}		\\
060729	& $	1.04	\pm{	0.03			}$ & $	0.86	\pm	0.02	$ &		&	\cite{Kann11}	\\
080319B	& $	2.30	^{+	0.70	}_{-	0.90	}$ & $	0.89	\pm	0.10	$ &		&	\cite{Bloom09}	\\
090618	& $	0.58	\pm{	0.05			}$ & $	0.99	\pm	0.05	$ &	uncorrected	&	This Work, [19]	\\
101225A	& $	0.08	\pm{	0.03			}$ & $	1.25	\pm	0.15	$ &		&	This Work	\\ \hline
\end{tabular}
\end{table*}

\section{Discussion on the redshift of GRB 101225A}

Determining the distance scale at which GRB\,101225A occurred is crucial to understand the energetics and get a clear picture of the physics involved in this event. In this section, we present several independent arguments to strengthen our redshift estimation.

The first strong limit on the redshift comes from the UVOT detection in $uvw2$ which implies a redshift lower than $z=1.4$ \cite{Campana10}.
As an independent test, the analysis of the absorption in the X-ray spectra (see Sect. 2 and in particular Fig.~\ref{fig:Xraycontours}) imposes an upper limit on the redshift of 0.5 within 99\% confidence and 0.35 within 90\% confidence.

The most restrictive redshift estimate comes from the SN fitting at 40 days after the burst (see Sect. 5.2), where we find a best fit for a broad-lined SN Ic, such as SN 1998bw, the prototype of GRB-related SNe, at $z=0.33^{+0.07}_{-0.04}$. For the rest of the SN types that give a reasonable fit, we obtain similar values, always resulting in a redshift between 0.21 and 0.50. Independently of the SN fit, we can compare the brightness of the bump in the light curve with the dimmest known SN (SN2008ha, which peaked at $M_R = -14.5\pm0.3$, \cite{Valenti09}) and take that as a lower limit on the redshift. In this case, we would expect GRB\,101225A to be located at a redshift larger than 0.1.

From the SED fit of the first days, we know that the evolution is well-described by a simple BB. Depending on the distance at which the object is found, we can derive different radii and expansion velocities. For an explosion of this type, we expect expansion velocities larger than $\sim10^3$ km s$^{-1}$, which would be at the edge of a stellar wind regime and, if similar to a SN explosion, of the order of $10^4$ km s$^{-1}$. We cannot, in principle, rule out higher velocities of the ejecta. Under the assumption that the ejecta should not be traveling at velocities larger than 100,000 km s$^{-1}$ ($0.3c$) and lower than 1,000 km s$^{-1}$ ($0.03c$), we can estimate a range of reasonable velocities between $z\sim0.20$ and $z\sim0.60$ (see Fig.~\ref{Fig:vel}). We note that there might be additional effects such as variations in the transparency of the BB that could introduce variations in this simplified analysis. In any case, the evolution of the BB would be hard to get by any source within the Local Group.
At a redshift of $z=0.33$, the velocity of the blackbody would have evolved from $\sim70,000$ km s$^{-1}$ at the time of our first SED to $\sim2,000$ km s$^{-1}$ at the time of our last blackbody-dominated epoch, nicely matching the requirements. The fact that the blackbody detected for XRF 060218/SN 2006aj was very similar to the one found for GRB\,101225A when placed at a redshift of $z=0.33$ (see Section \ref{sec:SNcomp}) adds additional evidence for the validity of the redshift estimate.

Concluding, the proposed redshift for GRB\,101225A of $z=0.33^{+0.07}_{-0.04}$ is supported by several independent arguments, and can be considered as a firm reference when studying the physical processes involved in the event. We do not find any evidence that would indicate a redshift smaller than $z=0.2$ or higher than $z=0.5$.

\section{Comparison between GRB\,101225A and other GRBs with SNe and BB components}
\label{sec:SNcomp}

In Table \ref{table:grbsne}, we compare several properties of GRBs associated with SNe and without a ``classical'' afterglow component. All of them are subluminous compared to the average long-duration GRB with E$_\mathrm{iso}$ around 10$^{51}-10^{54}$ erg. GRB 101225A lies on the lower end of the energy output from normal long-duration GRBs. Among those nearby GRB-SNe without a classical afterglow, there is a class of very long duration GRBs with very low E$_{peak}$ values, all of them showing a thermal component in X-rays. XRF 060218 and XRO 080109 also had a thermal component at optical wavelengths during the first few days [13, 22]. For XRF 100316D, no optical counterpart was detected before the onset of the actual SN due to high intrinsic extinction in the host galaxy.  

\begin{table*}[ht!]
 \caption{GRBs with SNe but without afterglows}
\label{table:grbsne}      
\begin{tabular}{l l l l l l l l l l}        
\hline                 
GRB       &      z   &            TC?     &        T$_{90}$&  E$_{peak}$       & E$_{iso}$   &   HR(50-100)   &   Radio? & SN M$_V$   & Host M$_B$\\
&&&(s)&(keV)&(erg)&&(mag)&(mag)\\ \hline
980425      &    0.0085 & No         &     23.3  &  55$\pm$21      &   8.1$\times 10^{47}$                  &  --- &            Yes        &     --19.42     &     --17.6\\
031203       &   0.105  & No          &    30       &       158$\pm$51   &     3$\times 10^{49}$                &   --- &                  Yes    &         --20.39       &   --21.0\\
060218      &    0.0331 & Yes       &      $\sim$2100  & 4.9                &     6.2$\times 10^{49}$          & 0.835 &          Yes       &      --18.76       &   --15.9\\
080109*    &     0.0065 & Yes      &       $\sim$400   & low                 &    2$\times 10^{46}$                 &     ---   &              Yes       &      --16.7       &    --20.7\\
100316D 	& 0.059  & Yes          &    $ >$1300          &  ---        &  3.1$\times 10^{49}$          & 0.891       &    No         &    --18.62          &             --18.8\\
101225A	& 0.33  & Yes      &       $>$2000   &38$\pm$20                 &     $>1.4 \times 10^{51}$      &     1.06      &      No        &      --16.9       &    --13.7\\ \hline
\vspace{0.5mm}
\end{tabular}

\footnotetext{1}{*No $\gamma$-rays observed, numbers derived from X-rays.}\\
\footnotetext{2}{TC? refers to the early thermal component, mostly attributed to a supernova breakout.}\\
\footnotetext{3}{HR is the hardness ratio, defined as the ratio of channels ($50-100$ kev)/($25-50$ keV)}\\
\footnotetext{4}{SN M$_V$ is the SN peak absolute magnitude in $V$}\\
\footnotetext{5}{Host M$_B$ is the host absolute magnitude in $B$}
\end{table*}

\begin{figure}[ht!]
   \centering
   \includegraphics[width=12cm]{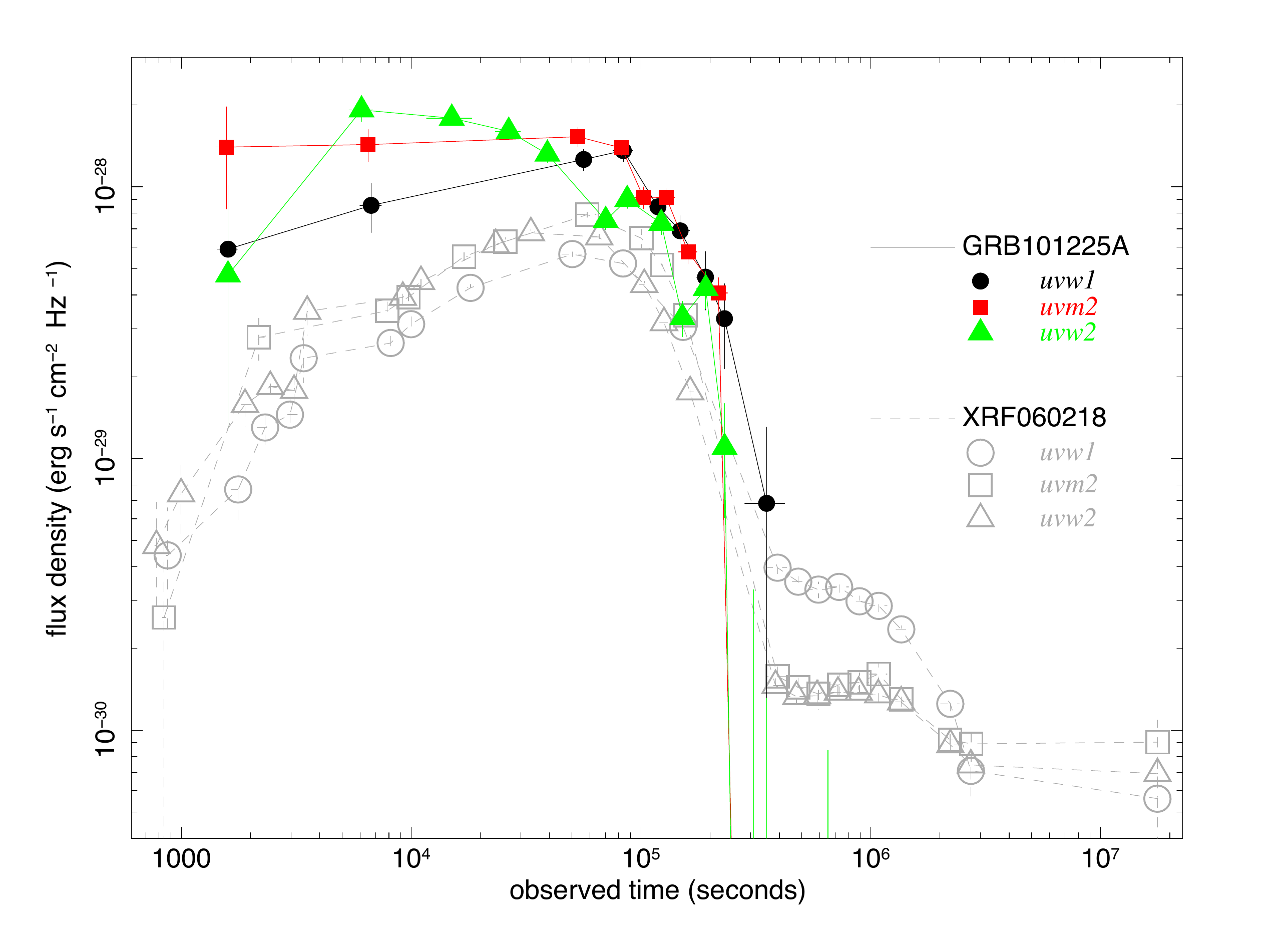}
      \caption{{\bf UV light curves in flux density for GRB 101225A and XRF 060218.} The light curves for XRF 060218 have been shifted to $z=0.33$ for direct comparison. The colored solid shapes connected by solid lines are for GRB 101225A, open symbols and dotted lines for XRF 060218. Circles are $uvw1$, squares are $uvm2$, triangles are $uvw2$.}
         \label{101225:UVOT}
   \end{figure}

XRF 060218 [13] shows a similar early behavior to GRB\,101225A. We compare the early UVOT light curve of XRF 060218, which we obtained from the UVOT catalogue \cite{Roming09}, to the light curve from GRB 101225A by shifting that of XRF 060218 to $z=0.3$, including a $K$-correction. To obtain the $K$-correction, we use {\tt XSPEC} assuming a blackbody spectrum with $\mathrm{kT}\sim3.7$ eV. The temperature was determined from the best-fit model of a BB to a SED of XRF 060218 taken at 120\,ks after the trigger [13]. Using this BB spectrum, we determined the expected flux density in the observed frame for each filter and at $z=0.33$. The ratio of these two flux densities was taken to be the $K$-correction for the specific filter, which is $\sim2.20$ for all filters. Figure \ref{101225:UVOT} compares the flux-density light curves of the 3 UV filters of both GRBs. For both GRBs the light curves were corrected for Galactic extinction.

\begin{figure}[!h]
      \centering
      \includegraphics[width=10.5cm]{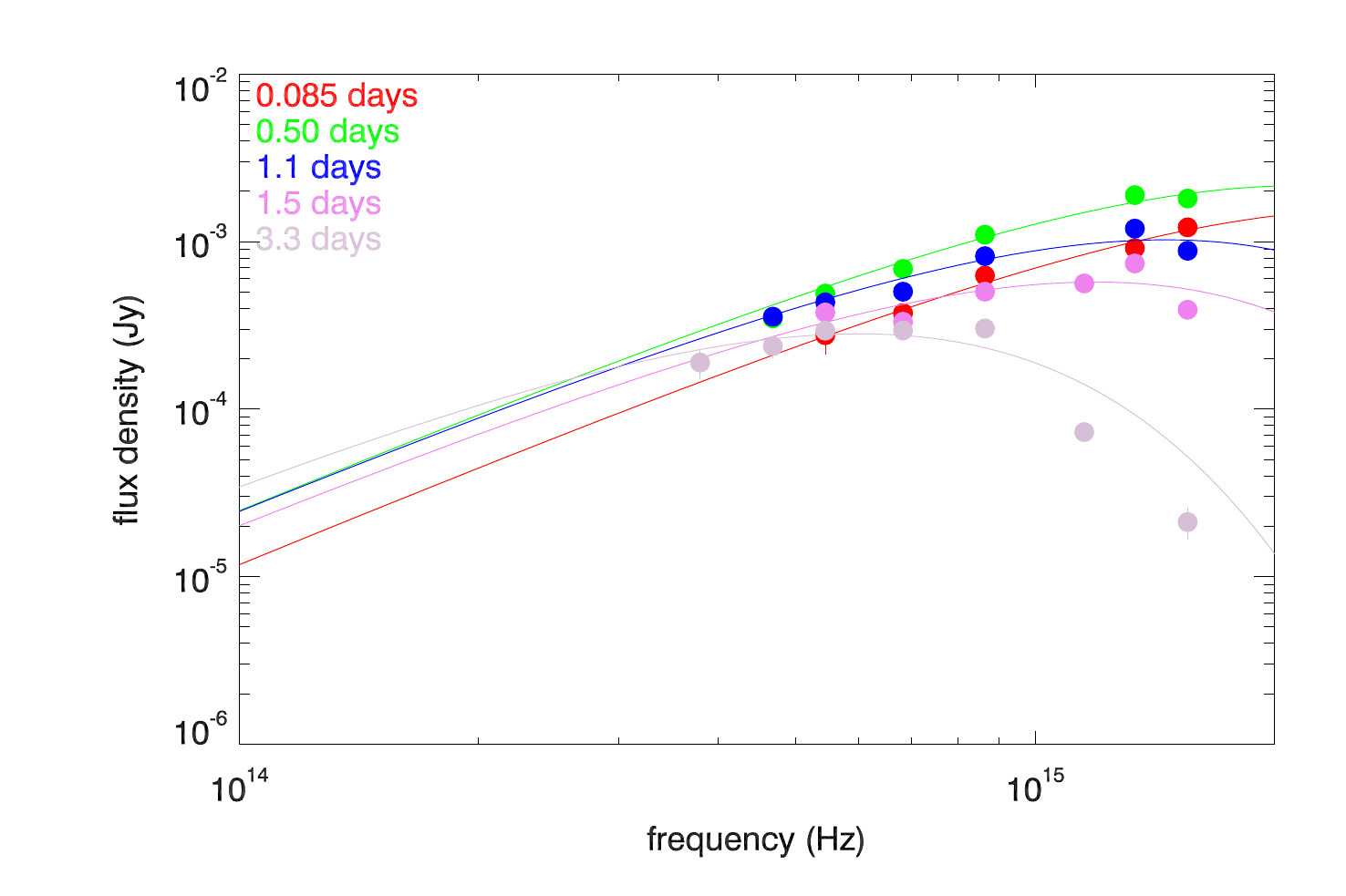}
          \includegraphics[width=10.5cm]{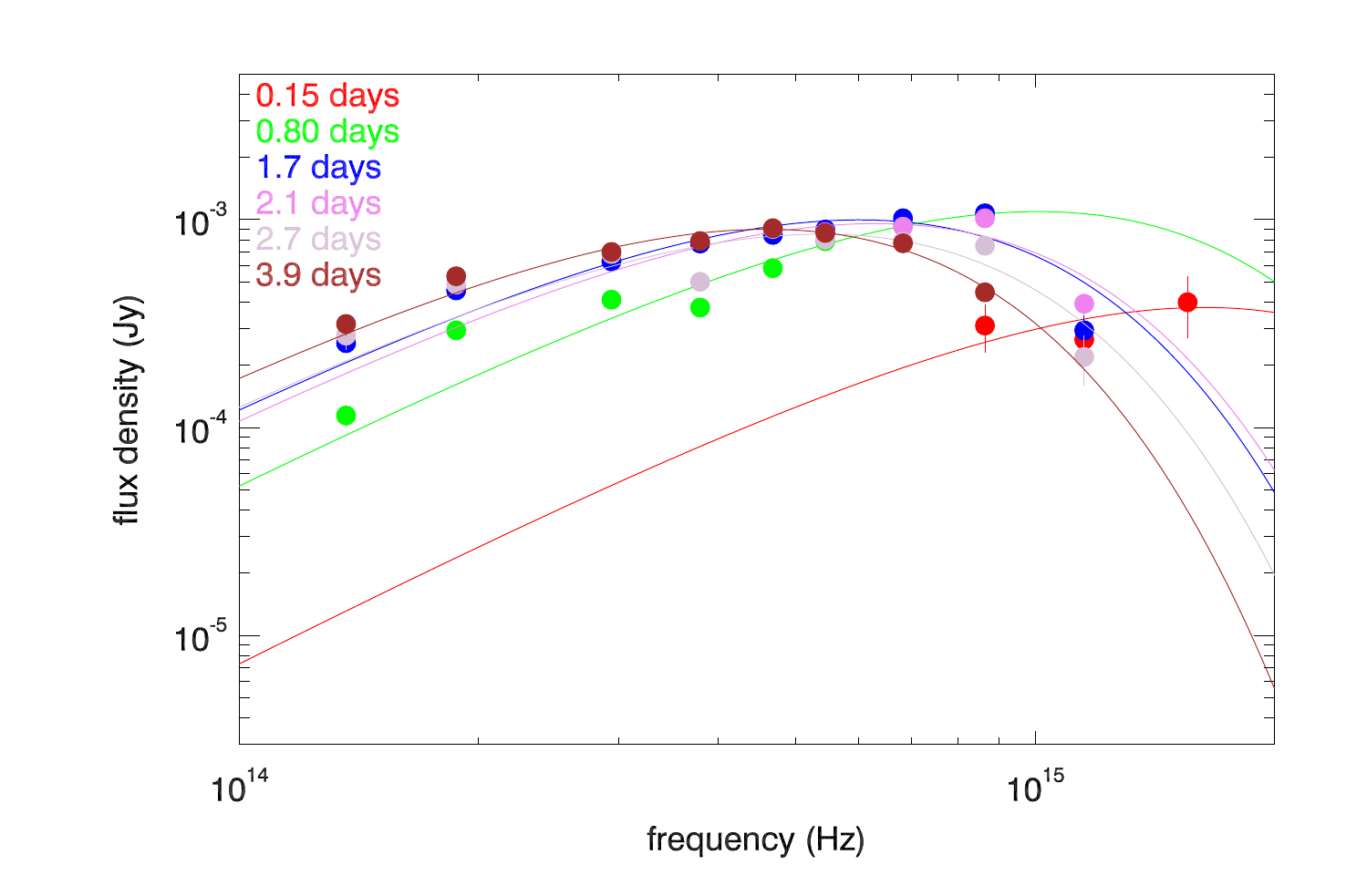}
      \caption{{\bf SED fit of the early UVOIR data of XRF 060218/SN 2006aj (top) and SN 2008D (bottom).} The last epoch in both plots already begin to show a strong contribution of the SN. These figures can be directly compared to Fig. 1 of the main paper.
              }
         \label{Fig:060218SED}
   \end{figure}

\begin{figure}[!h]
      \centering
      \includegraphics[width=9cm]{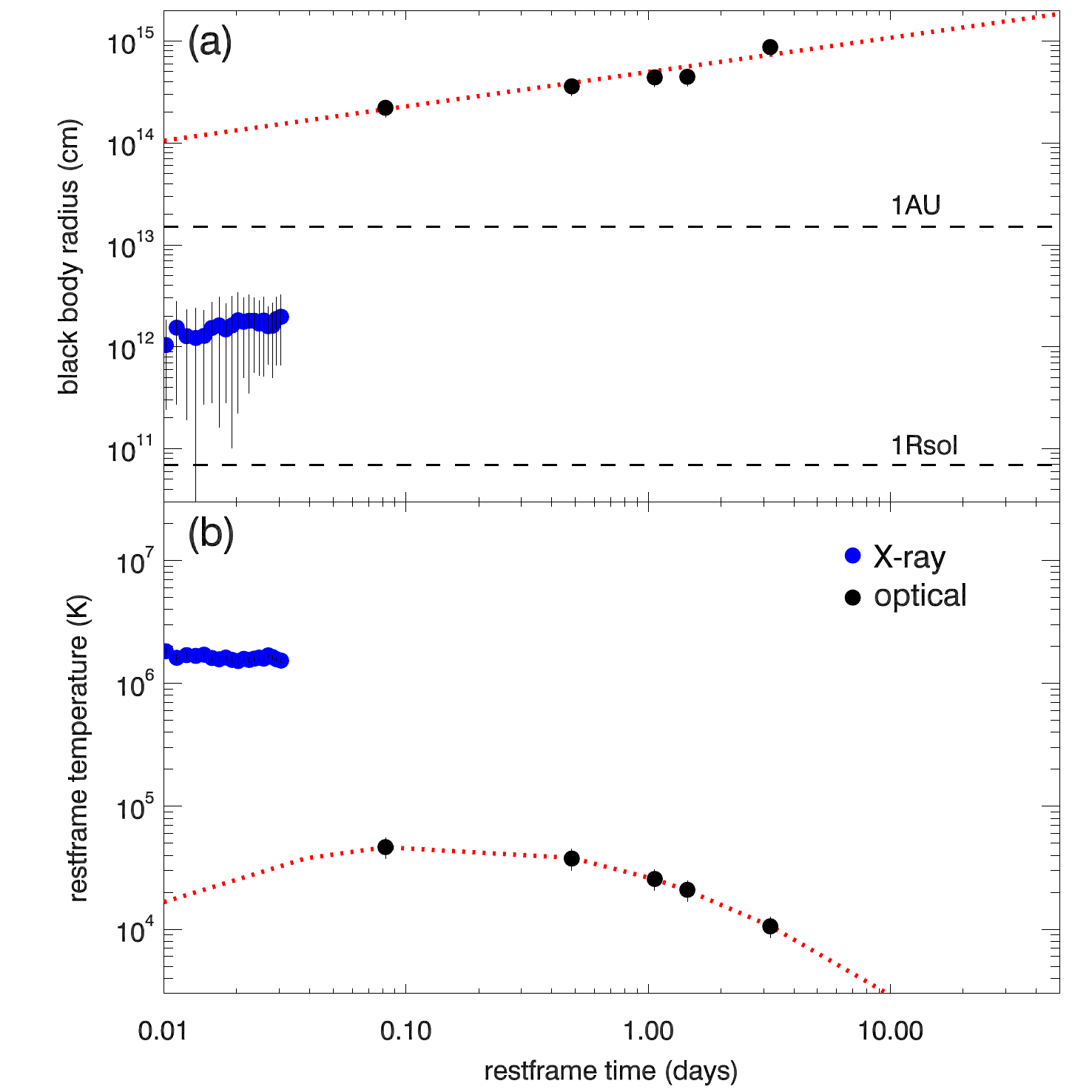}\\[6mm]
       \includegraphics[width=9cm]{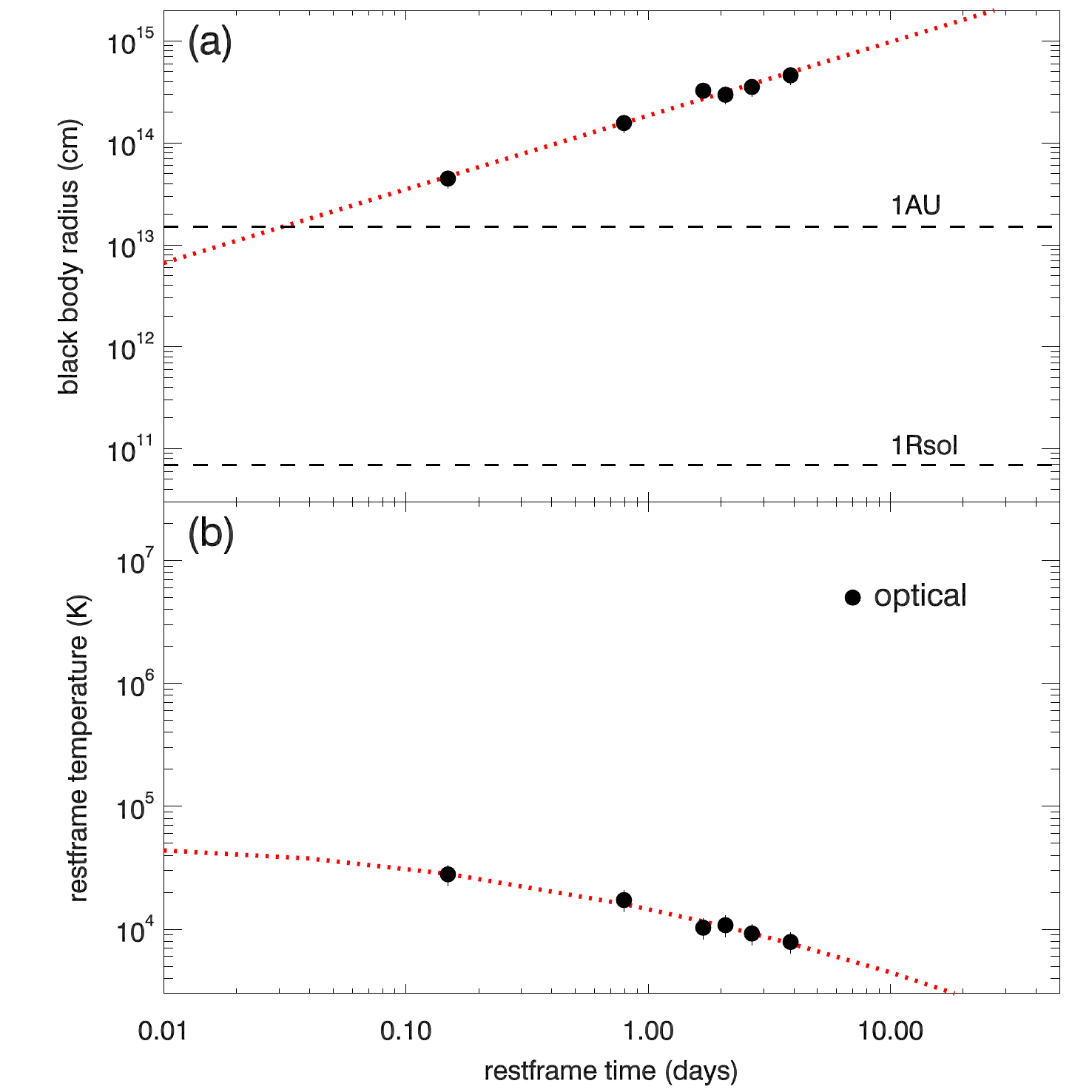}
      \caption{{\bf Temporal evolution of the radius and temperature for the pre-SN epochs of XRF 060218/SN 2006aj (top) and SN 2008D (bottom).} The dotted line shows a polynomial fit to the temperature and radius evolution. This figure can be directly compared to Fig. \ref{Fig:cte}. 
              }
         \label{Fig:060218rest}
   \end{figure}

We furthermore performed a fit of the early UVOIR SEDs of XRF\,060218 and SN 2008D, in a similar way to what we did for GRB\,101225A. XRF\,060218 also seems to follow a blackbody evolution at early times. However, the SN starts to dominate already around 3 days after the burst (see Fig. \ref{Fig:060218SED}), limiting how long the study of the evolution is possible. As can be seen from Fig. \ref{Fig:060218rest}, the evolution of the blackbody is not very different from what we see in GRB\,101225A, although the radius expansion is slightly steeper.
SN 2008D does not have a measurable thermal component in the early X-ray data. In the optical, the cooling BB dominates until about 4 days after the event when the onset of the SN was observed (see Fig.~\ref{Fig:060218SED}). The temperature of the UVOIR BB of SN 2008D is lower ($\sim$ 30 000 K) than those of XRF 060218 and GRB 101225A. The radius evolution is considerably steeper than for those two events (Fig.~\ref{Fig:060218rest}). The UVOIR BB emission of SN 2008D is therefore likely due to the cooling of the initial shock breakout.

\section{Temperature evolution and geometry of the UVOIR BB}{\label{sect:model}}

\begin{figure}[ht!]
      \centering
   \includegraphics[width=15cm]{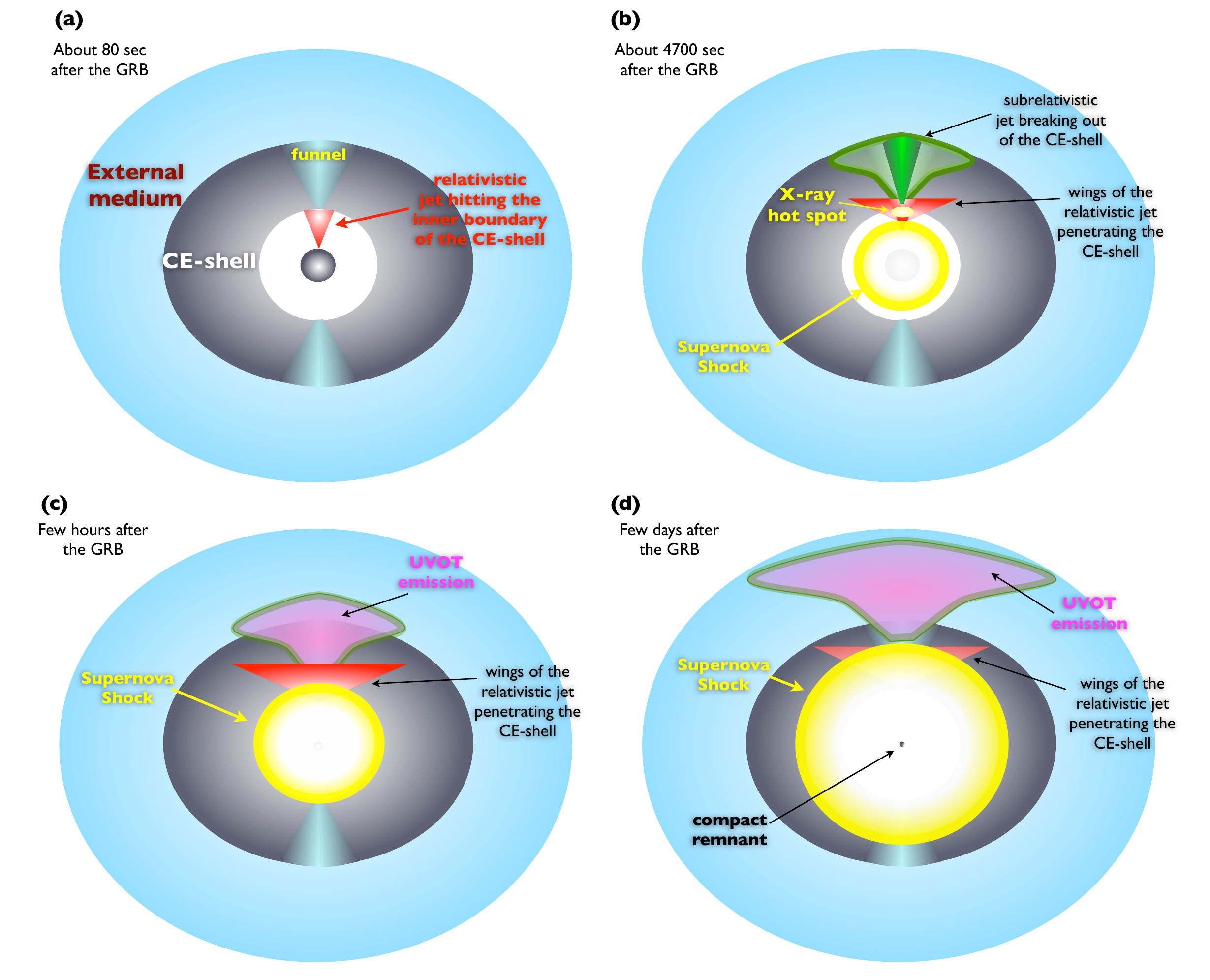}
      \caption{{\bf Cartoon of the different stages in our model of GRB 101225A.} The different stages are explained in the text.
              }
         \label{fig:cartoon}
   \end{figure}

The interaction of an ultra-relativistic, well-collimated jet (having a half-opening angle $\theta_{j,0}\sim
10^\circ$) with the previously ejected broad torus can explain the power-law and thermal component in the X-rays of GRB 101225A. In Fig. \ref{fig:cartoon} we show four different stages in the evolution of the jet and its interaction with the CE shell.

Taking the results from CE-simulations, we assume that the density of the common envelope ejecta in a narrow funnel around the symmetry axis of the system is much lower than elsewhere in the CE-ejecta. This funnel has an opening half-angle $\theta_f\sim 2^\circ$, which permits the passage of the $\gamma$-radiation generated by internal shocks in the ultra-relativistic jet. Since $\theta_f<\theta_{j,0}$, most of the jet beam hits the inner boundary of the CE-ejecta at a distance of $R_{CE,in}\simeq 2.5\times 10^{12}\,$cm (panel a of Fig. \ref{fig:cartoon}), while only a small fraction of the beam propagates through the funnel until it reaches the outer radial boundary of the CE-ejecta ($R_{CE,out}\simeq 2.1\times 10^{14}\,$cm, panel b).  As the central jet spine progresses through the ejecta funnel, it interacts with the lateral walls, giving rise to mass entrainment in the jet. The additional baryon load and the shear with the funnel walls decelerates the ultra-relativistic jet spine extremely quickly and, therefore, no regular afterglow signature is produced. Furthermore, most of the jet beam hits regions of the CE-ejecta which are much denser than the central funnel and, thus, its propagation across the ejecta is much slower than that of the central spine.  As the jet decelerates to moderately relativistic speeds it expands laterally.

The X-ray emission is attributed to the shocks (forward and reverse) produced as the ultra-relativistic jet impinges against the inner radial boundary of the CE-ejecta. In our model the cross-sectional radius of the CE-ejecta funnel ($R_f=R_{CE,in} \sin{\theta_f}\sim 10^{11}\,$cm) sets the almost-constant size of the observed X-ray source (blue dots in Fig. \ref{Fig:cte}). 

An approximate evolution of the jet spine as it crosses the ejecta funnel can be obtained by applying the model of \cite{Huang00}, which is based on solving a system of four ordinary differential equations (as a function of the observer's time $t$) for the Lorentz factor, the rest-mass, the radius and the jet half-opening angle, respectively:

\begin{eqnarray}
\frac{d\Gamma}{dt} &=& \frac{dm}{dt} \frac{1-\Gamma^2}{M_{jet}+\epsilon
  m + 2(1-\epsilon)\Gamma m} \label{eq:gammaevol} \\
 \frac{dm}{dt} &=& \frac{dR}{dt} 2\pi R^2 (1-\cos{\theta}) n(R) \\ 
\frac{dR}{dt} &=& \beta c \Gamma (\Gamma + \sqrt{\Gamma^2-1}) \\
\frac{d\theta}{dt} &=& \frac{c_s(\Gamma + \sqrt{\Gamma^2-1})}{R},
\label{eq:thetaevol}
\end{eqnarray}
where $n(R)$ is the particle number density, $\beta=\sqrt{1-\Gamma^{-2}}$ is the velocity, $\epsilon$ is the shock-generated thermal energy (in the comoving frame) that is
radiated, and $c_s=\sqrt{(4\Gamma+1)(\Gamma^2-1)(3\Gamma(4\Gamma^2-1))^{-1}}$ is the sound speed. The temperature of the jet spine is derived by assuming that the pressure is dominated by the photon radiation, so that $P(t) = aT^4(t)/3$, with the radiation constant $a = 7.56\times 10^{-15}$\,erg\,cm$^{-3}$K$^{-4}$, and noting that $P=(\Gamma^2-1)/(3\Gamma)n m_p c^2$, ($m_p$ being the proton mass) according to \cite{Huang00}.

For the jet spine propagating through the CE-ejecta funnel we may take the following initial conditions, corresponding to an ultra-relativistic and cold jet: $\Gamma_{in}\sim 100$, $M_{in} \simeq 8\times10^{-9}M_\odot$, $R_{in}=R_{CE,in}=2.5\times 10^{12}\,$cm, and $\theta_{in}=\theta_f=2^\circ$. A moderate radiative efficiency $\epsilon\simeq 0.4$ is assumed. The initial jet mass is estimated from the observed lower limit of $E_{\gamma,iso}$ and the initial jet half-opening angle as $M_{jet}=E_{jet}/(((\Gamma-1)(1-\epsilon) + \epsilon)c^2)$, where $E_{jet} = E_{\gamma,iso}(1-\cos{\theta_f}) \simeq 8.5 \times 10^{47}\,$erg. We assume that the CE-ejecta have a declining particle number density of the form $n(R)=n_{in}(R_{CE,in}/R)^2$, with $n_{in}=4.9\times 10^{12}\,$cm$^{-3}$. After $t\simeq 0.05\,$days the jet spine reaches $R_{CE,out}$ at a velocity of $v_{jet,out}\sim 0.25c$, an opening half-angle of $\theta_{jet,out}\simeq 70^\circ$, and having plowed $\simeq 1.5\times 10^{-5}M_\odot$ of the CE-ejecta.

\begin{figure}[!ht]
   \centering
   \includegraphics[width=11cm]{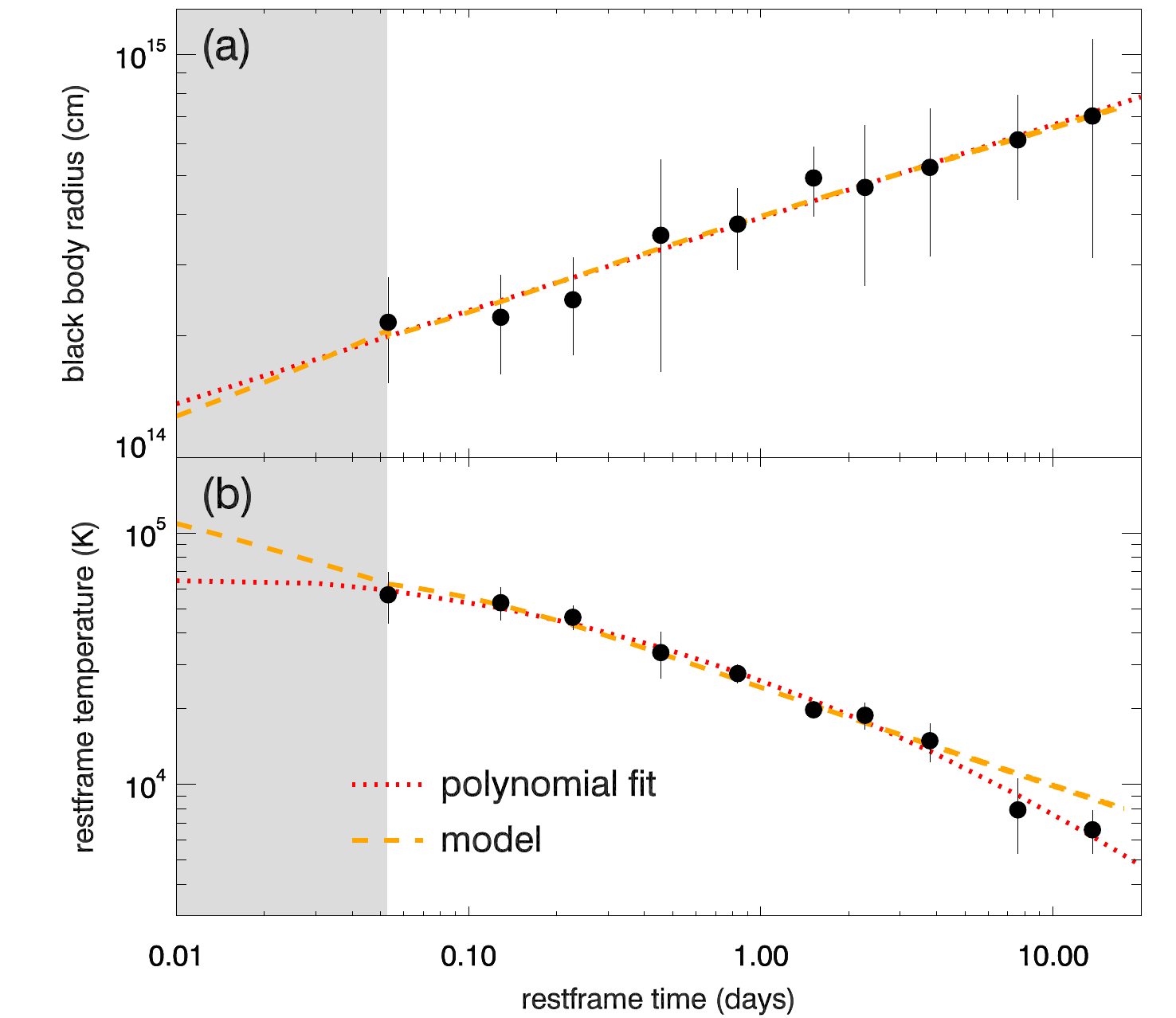}
   \caption{{\bf Radius (panel a) and temperature (panel b) evolution of a sub-relativistic jet}. The initial evolution up to $\simeq 0.05\,$days (grey shaded area), is computed with a modified version of the
    \cite{Huang00} method to incorporate a shell profile in
     which the number density decays with radius as
     $n=n_{in}(R_{CE,in}/R)^2$, with $n_{in}=4.9\times
     10^{12}\,$cm$^{-3}$. For this part of the evolution, the initial
     data are: $R_{in}=R_{CE,in}=2.5\times 10^{12}\,$cm, $E_{jet}\simeq 8.5\times
     10^{47}\,$erg, $\Gamma_{in}\sim 100$,
     $\theta_{in}=\theta_f=2^\circ$, and $\epsilon\simeq 0.4$. }
         \label{Fig:Tevoltheory}
\end{figure}

The temperature and radius evolution of the UVOIR BB (panel c in Fig. \ref{fig:cartoon}) can be modeled as the result of the further deceleration of the sub-relativistic jet emerging trough $R_{CE,out}$. Again, we use equations \ref{eq:gammaevol}--\ref{eq:thetaevol}, but now considering a uniform medium with a particle density of $n_{ext}\simeq 1.6\times 10^9\,$cm$^{-3}$ and a higher radiative efficiency of $\epsilon=1$. As initial conditions we take the terminal values of the velocity, mass, radius and opening half-angle of the previous evolution through the CE-ejecta funnel. The result can be seen in Fig.~\ref{Fig:Tevoltheory} (after $t\simeq 0.05\,$days). Small variations of the initial variables can also roughly fit both the radius and the temperature evolution. We note that, even though the theoretical model does not predict an exact power-law for the radius evolution, the deviations from the observed $R(t)\propto t^{0.22}$ are rather small. We also point out that the temperature evolution is not compatible with a single power-law as the observational data suggests.

Finally, we speculate on the fate of the jet wings, i.e., the fraction of the jet that does not cross the CE-ejecta funnel. The energy contained in these wings, $\sim 10^{49}\,$erg, is transferred to the CE-ejecta, and can break out through $R_{CE,out}$ on scales of $\sim\,$few days, almost isotropically, but at relatively low temperatures compared with the emerging jet spine. Hence, its observational signature will be probably hidden by the emerging SN light curve (panel d) in Fig. \ref{fig:cartoon}.

\end{document}